\documentclass[preprintnumbers,nofootinbib,superscriptaddress,10pt,secnumarabic,aps,pra]{revtex4-1}

\usepackage{amsfonts,amssymb,amsmath,amsthm,graphicx}
\usepackage{url}
\usepackage{dsfont}
\usepackage{bbm}

\theoremstyle{definition}
\newtheorem{theorem}{Theorem}
\newtheorem{proposition}[theorem]{Proposition}

\newtheorem{lemma}[theorem]{Lemma}
\newtheorem{definition}[theorem]{Definition} 
\newtheorem{corollary}[theorem]{Corollary}
\newtheorem{remark}[theorem]{Remark}

\newcommand*{\cB}{\mathcal{B}}

\newcommand*{\cD}{\mathcal{D}}
\newcommand*{\cE}{\mathcal{E}}
\newcommand*{\cF}{\mathcal{F}}

\newcommand*{\cH}{\mathcal{H}}
\newcommand*{\cI}{\mathcal{I}}

\newcommand*{\cL}{\mathcal{L}}

\newcommand*{\cN}{\mathcal{N}}

\newcommand*{\cP}{\mathcal{P}}
\newcommand*{\cS}{\mathcal{S}}

\newcommand*{\cT}{\mathcal{T}}
\newcommand*{\cV}{\mathcal{V}}

\newcommand*{\cX}{\mathcal{X}}
\newcommand*{\cY}{\mathcal{Y}}

\newcommand*{\tr}{\mathrm{tr}}
\newcommand*{\ket}[1]{| #1 \rangle}
\newcommand*{\bra}[1]{\langle #1 |}

\newcommand*{\proj}[1]{\ket{#1}\bra{#1}}

\newcommand*{\genFid}{\bar{F}}
\newcommand*{\eps}{\varepsilon}

\newcommand*{\1}{\mathbbm{1}}

\newcommand{\mycomment}[1]{}

\begin{document}

\title{Entanglement Cost of Quantum Channels}

\author{Mario \surname{Berta}}
\email[]{berta@caltech.edu}
\affiliation{Institute for Theoretical Physics, ETH Zurich, 8093 Zurich, Switzerland.}
\author{Fernando G.S.L. \surname{Brand\~{a}o}}
\email[]{fgslbrandao@gmail.com}
\affiliation{Institute for Theoretical Physics, ETH Zurich, 8093 Zurich, Switzerland.}
\affiliation{Departamento de Fisica, Universidade Federal de Minas Gerais, Belo Horizonte 30123-970, Brazil.}
\affiliation{Centre for Quantum Technologies, National University of Singapore, 2 Science Drive 3, 117543 Singapore.}
\author{Matthias \surname{Christandl}}
\email[]{christandl@math.ku.dk}
\affiliation{Institute for Theoretical Physics, ETH Zurich, 8093 Zurich, Switzerland.}
\author{Stephanie \surname{Wehner}}
\email[]{s.d.c.wehner@tudelft.nl}
\affiliation{Centre for Quantum Technologies, National University of Singapore, 2 Science Drive 3, 117543 Singapore.}

\date{\today}


\begin{abstract}
The entanglement cost of a quantum channel is the minimal rate at which entanglement (between sender and receiver) is needed in order to simulate many copies of a quantum channel in the presence of free classical communication. In this paper we show how to express this quantity as a regularised optimization of the entanglement formation over states that can be generated between sender and receiver. Our formula is the channel analog of a well-known formula for the entanglement cost of quantum states in terms of the entanglement of formation; and shares a similar relation to the recently shattered hope for additivity.

The entanglement cost of a quantum channel can be seen as the analog of the quantum reverse Shannon theorem in the case where free classical communication is allowed. The techniques used in the proof of our result are then also inspired by a recent proof of the quantum reverse Shannon theorem and feature the one-shot formalism for quantum information theory, the post-selection technique for quantum channels as well as Sion's minimax theorem.

We discuss two applications of our result. First, we are able to link the security in the noisy-storage model to a problem of sending quantum rather than classical information through the adversary's storage device. This not only improves the range of parameters where security can be shown, but also allows us to prove security for storage devices for which no results were known before. Second, our result has consequences for the study of the strong converse quantum capacity. Here, we show that any coding scheme that sends quantum information through a quantum channel at a rate larger than the entanglement cost of the channel has an exponentially small fidelity.
\end{abstract}

\maketitle


\section{Introduction}

The quantification of the information theoretic power of quantum channels is one of the most fundamental problems in quantum information theory. Of particular interest is thereby the study of a channel's capacity for information transmission. This quantity corresponds to the number of bits $m$ that can be sent reliably when using the channel $n$ times using optimal encoding and decoding operations. Unlike classical channels, quantum channels have various distinct capacities, depending on the kind of information that is sent (e.g.~classical or quantum) or on the kind of assistance that is allowed (e.g.~free entanglement or free classical communication). Important examples of quantum channel capacities include the entanglement assisted classical capacity $C_{E}$~\cite{Bennett02}, and the classical communication assisted quantum capacities $Q_{\rightarrow}$, $Q_{\leftarrow}$ and $Q_{\leftrightarrow}$ depending on the direction of the assisting communication~\cite{Devetak_05,Shor_02,Lloyd97}.\\

One way of tackling the problem of capacities is to think more broadly in terms of channel simulations. For example, the process of sending $m$ bits reliably using $n$ uses of a channel $\cE$ can be understood as a simulation of $m$ perfect, noise-free, channels using $n$ copies of $\cE$. The capacity of the channel $\cE$ is then simply the rate $m/n$ at which such a simulation is possible in the limit of large $n$. One can also turn the problem upside down and ask: What is the optimal rate at which a perfect channel can simulate a noisy one? When the simulation can consume free entanglement between the sender and the receiver, this question is answered by the quantum reverse Shannon theorem. It states that the optimal rate is given by the entanglement assisted classical capacity $C_{E}$~\cite{Bennett06,Berta09_2}. Apart from its deep conceptual appeal, the quantum reverse Shannon theorem led to the proof that the $C_{E}$ is in fact a strong converse capacity.\\

It is natural to ask how these capacities change in the presence of other free resources. In this work, we consider the simulation of a noisy quantum channel $\cE$ by a noise-free channel in the presence of free classical communication. It turns out not to matter whether we allow free classical forward, backward, or even two-way communication, the capacity is the same in all scenarios. The problem we are considering can therefore be understood as the \lq reverse problem\rq~for all three classical communication assisted quantum capacities. Note that by quantum teleportation~\cite{teleport}, the perfect quantum channel can equivalently be replaced with perfect entanglement. The central question of this paper can thus be summarized as
\begin{center}
\begin{minipage}{0.62\textwidth}
{\it At what rate is entanglement, in the form of ebits, needed in order to asymptotically simulate a quantum channel $\cE$, when classical communication is given for free? }
\end{minipage}
\end{center}
We call this rate the entanglement cost $E_C$ of a quantum channel. Our main contribution in this paper is to prove the following formula
\begin{align}\label{eq:first}
E_{C}(\cE)=\lim_{n\rightarrow\infty}\frac{1}{n}\max_{\psi^{n}}E_{F}\left(\left(\cE^{\otimes n}\otimes\cI\right)\left(\psi^{n}\right)\right)\ ,
\end{align}
where the maximization is over all purifications $\psi^{n}$ of input states to the $n$-fold tensor product quantum channel $\cE^{\otimes n}$ and $\cI$ stands for the identity channel on the purifying system. The entanglement of formation $E_{F}$ is computed between purifying system and channel output; it is defined as
\begin{align}
E_{F}(\rho_{AB})=\inf_{\{p_{i},\rho^{i}\}}\sum_{i}p_{i}H(A)_{\rho^{i}}\ ,
\end{align}
where the infimum ranges over all pure state decompositions $\rho_{AB}=\sum_{i}p_{i}\rho^{i}_{AB}$, and $H(\cdot)$ denotes the von Neumann entropy. Note that expression~\eqref{eq:first} involves a regularization, and is therefore not a single-letter formula. Even if we would know that we can restrict the maximization to non-entangled input states, equation~\eqref{eq:first} would still not reduce to such a formula, due to Hasting's counterexample for the additivity of the entanglement of formation~\cite{Hastings09,Shor04}.\footnote{However we want to emphasize that we can compute explicit upper bounds for $E_{C}$, which are particularly useful for the applications given below.} Note also that $E_{C}$ is generally larger than $Q_\leftrightarrow$,\footnote{The same applies to $Q_\rightarrow$ and $Q_\leftarrow$ since both are smaller or equal to $Q_\leftrightarrow$.} in fact more strikingly, there exist so-called bound entangled channels $\cE$ (for instance entangling PPT channels) for which $E_C(\cE)>Q_\leftrightarrow(\cE)=0$. This fact highlights an important difference compared to the case of free entanglement where the quantum reverse Shannon theorem implies that the corresponding rates are equal. In particular, when $E_{C}(\cE)>Q_\leftrightarrow(\cE)$, the concatenated protocol which first simulates $\cE$ from a noiseless channel and then the noiseless channel from $\cE$ will result in a net loss.\\

As the name entanglement cost suggests, $E_{C}(\cE)$ is the quantum channel analog of the entanglement cost of quantum states $E_{C}(\rho_{AB})$, which corresponds to the rate of entanglement needed in order to generate a bipartite quantum state $\rho_{AB}$~\cite{Bennett96}. Our formula \eqref{eq:first} can be seen as the channel analog of the following well-known result for the quantum state problem~\cite{Hayden01}
\begin{align}
E_{C}(\rho_{AB})=\lim_{n\rightarrow\infty}\frac{1}{n}E_{F}(\rho_{AB}^{\otimes n})\ .
\end{align}
and the gap between $E_C(\cE)$ and $Q_\leftrightarrow(\cE)$ has its analog in the gap between $E_C(\rho_{AB})$ and $E_D(\rho_{AB})$, the distillable entanglement.\\

We present two applications. The first one concerns the security in the noisy-storage model~\cite{Noisy1,Wullschleger09,serge:bounded}. For the first time, we relate security in this model to a problem of sending quantum rather than classical information through the adversary's storage device. In particular, we show that any two-party cryptographic primitive can be implemented securely whenever
\begin{align}
E_C(\cE) \cdot \nu < \frac{1}{2}\ ,
\end{align}
where the adversary's storage is of the form $\cE^{\otimes \nu \cdot m}$, $m$ is the number of qubits transmitted during the protocol, and $\nu$ is the storage rate (see Section~\ref{sec:noisy} for precise definitions). Our analysis improves the range of parameters when security can be obtained. We illustrate our results with explicit calculations for a number of specific channels. In particular, we obtain non-trivial bounds for dephasing noise and for any qubit channel - for instance the amplitude damping channel.

The second application of our result is an upper bound on the strong converse capacity for sending quantum information. The strong converse capacity is the minimal rate above which any attempt to send information necessarily has exponentially small fidelity.\footnote{Note that the strong converse capacity is greater or equal than the standard capacity (which is defined as the minimal rate above which the fidelity does not approach one).} The strong converse capacity for sending classical information is known to be equal to the classical capacity for a selected number of channels~\cite{Wehner_K09}, or under additional assumptions~\cite{winter99,ogawa99}. For many channels there are also upper bounds known~\cite{Bennett06,fernando1,ciara:converse}, but a general formula for the strong converse classical capacity is not known. Understanding the strong converse capacity for sending quantum information turns out to be an even more elusive problem, and the only previous result relies on a statement involving the transmission of classical information~\cite{Bennett06}. Here, we make progress by showing that any coding scheme sending quantum information (using free forward, backward or two-way classical communication) at an asymptotic rate higher than the entanglement cost $E_{C}$, must have an exponentially small fidelity.\\

The proof of our main result~\eqref{eq:first} is based on one-shot information theory, which makes statements about structureless resources avoiding the usual requirement of independence and identical distribution (i.i.d.). The role of von Neumann entropies in the i.i.d.~scenario is taken by min- and max-entropies from the smooth entropy formalism~\cite{Ren05,wolf:04n,renner:04n,koenig-2008,Tomamichel08,Tomamichel09,datta-2008-2}. We work in this formalism and the proof of our main result is conceptually very similar to the proof of the quantum reverse Shannon theorem given in~\cite{Berta09_2}. In order to prove the direct part of~\eqref{eq:first}, we need to show the existence of a channel simulation for $\cE^{\otimes n}$, whose asymptotic rate of entanglement consumption is upper bounded by $E_{C}(\cE)$. That is, we need to construct a completely positive and trace preserving (CPTP) map that is arbitrarily close to $\cE^{\otimes n}$ in the diamond norm\footnote{The diamond norm is the dual of the completely bounded norm~\cite{Kitaev97}.} and that uses local operations and classical communication as well as ebits at a rate of at most $E_{C}(\cE)$. Here it is worth noting that even though the channel we wish to simulate has i.i.d.~structure, the channel simulation also has to work on non-i.i.d.~inputs. The crucial idea in order to deal with this fact is to employ the post-selection technique for quantum channels~\cite{ChristKoenRennerPostSelect}, which is a tool to bound the distance in diamond norm between two completely positive and trace preserving (CPTP) maps. The technique upper bounds this distance by the distance arising from the purification of a special de Finetti input state.\footnote{A de Finetti state consists of n identical and independent copies of an (unknown) state on a single subsystem.} With this, it is sufficient to find a CPTP map that does the channel simulation on the purification of this special de Finetti state, and to quantify how much entanglement this consumes. Since the state is a purification of a de Finetti state (and not a de Finetti state itself) it does not have i.i.d.~structure. In order to deal with this fact we employ ideas from the one-shot entanglement cost for quantum states $E_{C}^{(1)}(\rho_{AB},\eps)$, which quantifies how much entanglement is needed in order to create one single copy of a bipartite quantum state $\rho_{AB}$ using local operations and classical communication~\cite{datta-2008-7,HayashiBook}.\footnote{This is in contrast to the quantity $E_{C}(\rho_{AB})$ mentioned before, which answers the question of how much entanglement is needed in the asymptotic i.i.d.~regime.} The resulting entanglement cost of the channel simulation is then upper bounded by an expression similar to~\eqref{eq:first}, but with the maximization over input states and the minimization in the definition of the entanglement of formation interchanged. Finally, in order to arrive at~\eqref{eq:first}, we discretize the set of Kraus decompositions of $\cE$ and apply von Sion's minimax theorem to swap the minimization and the maximization~\cite{Sion58}. The proof of the converse follows a standard argument applied to the one-shot entanglement cost.\\

This paper is structured as follows. In Section~\ref{sec:pre} we introduce notation, definitions and state some basic lemmas. In particular, we review the results of~\cite{datta-2008-7} about the one-shot entanglement cost of quantum states. In Section~\ref{sec:main} we derive our main result; we define and quantify the entanglement cost of quantum channels. This is followed by a discussion of applications in Section~\ref{sec:app}. Finally we end with a summary and give an outlook (Section~\ref{sec:dis}). The arguments are based on various technical statements, which are proven in Appendices~\ref{app:smooth} -~\ref{app:tech}.


\section{Preliminaries}\label{sec:pre}

We assume that all Hilbert spaces, in the following denoted $\cH$, are finite-dimensional. The dimension of $\cH_{A}$ is denoted by $|A|$. The set of linear operators on $\cH$ is denoted by $\cL(\cH)$ and the set of positive semi-definite operators on $\cH$ is denoted by $\cP(\cH)$. We define the sets of sub-normalized states $\cS_{\leq}(\cH)=\{\rho\in\cP(\cH):\tr[\rho]\leq1\}$, normalized states $\cS(\cH)=\{\rho\in\cP(\cH):\tr[\rho]=1\})$, and normalized pure states $\cV(\cH)=\{\proj{\psi}\in\cS(\cH):\ket{\psi}\in\cH\}$. The tensor product of $\cH_{A}$ and $\cH_{B}$ is denoted by $\cH_{AB}\equiv\cH_{A}\otimes\cH_{B}$. Given a multipartite operator $\rho_{AB}\in\cP(\cH_{AB})$, we write $\rho_{A}=\tr_{B}[\rho_{AB}]$ for the corresponding reduced operator. For $M_{A}\in\cL(\cH_{A})$, we write $M_{A}\equiv M_{A}\otimes\1_{B}$ for the enlargement on any $\cH_{AB}$, where $\1_{B}$ denotes the identity in $\cL(\cH_{B})$. For $\cH_{A}$, $\cH_{B}$ with orthonormal bases $\{\ket{i}_{A}\}_{i=1}^{|A|}$, $\{\ket{i}_{B}\}_{i=1}^{|B|}$ and $|A|=|B|$, the canonical identity mapping from $\cL(\cH_{A})$ to $\cL(\cH_{B})$ with respect to these bases is denoted by $\cI_{A\rightarrow B}$, i.e.~$\cI_{A\rightarrow B}(\ket{i}\bra{j}_{A})=\ket{i}\bra{j}_{B}$. A linear map $\cE_{A\rightarrow B}:\cL(\cH_{A})\rightarrow\cL(\cH_{B})$ is positive if $\cE_{A\rightarrow B}(\rho_{A})\in\cP(\cH_{B})$ for all $\rho_{A}\in\cP(\cH_{A})$. It is completely positive if the map $(\cE_{A\rightarrow B}\otimes\cI_{C\rightarrow C})$ is positive for all $\cH_{C}$. Completely positive and trace preserving maps are called CPTP maps or quantum channels. The support of $\rho\in\cP(\cH)$ is denoted by $\mathrm{supp}(\rho)$, the projector onto $\mathrm{supp}(\rho)$ is denoted by $\rho^{0}$ and $\tr\left[\rho^{0}\right]=\mathrm{rank}(\rho)$, the rank of $\rho$. For $\rho\in\cP(\cH)$ we write $\|\rho\|_{\infty}$ for the operator norm of $\rho$, which is equal to the maximum eigenvalue of $\rho$. The trace norm of $M\in\cL(\cH)$ is defined as $\|M\|_{1}=\tr\left[\sqrt{M^{\dagger}M}\right]$, and the Hilbert-Schmidt norm of $M$ is given by $\|M\|_{2}=\sqrt{\tr\left[M^{\dagger}M\right]}$.\\

Recall the following standard definitions. The \textit{von Neumann entropy} of $\rho\in\cP(\cH)$ is defined as $H(\rho)=-\tr[\rho\log\rho]$,\footnote{$\log$ denotes the logarithm to base 2.} and the \textit{conditional von Neumann entropy} of $A$ given $B$ for $\rho_{AB}\in\cP(\cH_{AB})$ is given by
\begin{align}
H(A|B)_{\rho}=H(AB)_{\rho}-H(B)_{\rho}\ .
\end{align}

\begin{definition}\label{def:formation}
Let $\rho_{AB}\in\cS(\cH_{AB})$. The \textit{entanglement of formation} of $\rho_{AB}$ is defined as
\begin{align}
E_{F}(\rho_{AB})=\inf_{\{p_{i},\rho^{i}\}}\sum_{i}p_{i}H(A)_{\rho^{i}}=\inf_{\{p_{i},\rho^{i}\}}H(A|R)_{\rho}\ ,
\end{align}
where the infimum ranges over all pure states decompositions $\rho_{AB}=\sum_{i}p_{i}\rho_{AB}^{i}$ and $\rho_{AR}=\sum_{i}p_{i}\rho^{i}_{A}\otimes\ket{i}\bra{i}_{R}$.
\end{definition}

\begin{definition}\label{def:altmax}
Let $\rho_{AB}\in\cP(\cH_{AB})$. The \textit{alternative max-entropy} of $A$ conditioned on $B$ is defined as
\begin{align}
H_{0}(A|B)_{\rho}=\sup_{\sigma_{B}\in\cS(\cH_{B})}\log\tr\left[\rho_{AB}^{0}(\1_{A}\otimes\sigma_{B})\right]\ .
\end{align}
\end{definition}

In the literature this quantity is also known as conditional max-entropy~\cite{Ren05,datta-2008-2} or conditional zero-R\'enyi entropy~\cite{datta-2008-7}. We will evaluate the alternative conditional max-entropy in particular on quantum-classical states.

\begin{lemma}\label{lem:h0class}
Let $\rho_{AB}\in\cP(\cH_{AB})$ with $\rho_{AB}=\sum_{k\in K}\rho_{A}^{k}\otimes\proj{k}_{B}$, $\rho_{A}^{k}\in\cP(\cH_{A})$, and the $\ket{k}_{B}$ mutually orthogonal (i.e.~the state is classical on $B$). Then,
\begin{align}
H_{0}(A|B)_{\rho}=\max_{k\in K}H_{0}(A)_{\rho^{k}}\ .
\end{align}
\end{lemma}

\begin{proof}
We calculate
\begin{align}
H_{0}(A|B)_{\rho}&=\max_{\sigma_{B}\in\cS(\cH_{B})}\log\tr\left[\rho_{AB}^{0}(\1_{A}\otimes\sigma_{B})\right]\\
&=\log\max_{\sigma_{B}\in\cS(\cH_{B})}\tr\left[\left(\sum_{k}\left(\rho_{A}^{k}\right)^{0}\otimes\proj{k}_{B}\right)\left(\1_{A}\otimes\sigma_{B}\right)\right]\\
&=\log\max_{\sigma_{B}\in\cS(\cH_{B})}\tr\left[\sigma_{B}\cdot\left(\sum_{k}\proj{k}_{B}\cdot\tr\left[\left(\rho_{A}^{k}\right)^{0}\right]\right)\right]\\
&=\log\left\|\sum_{k}\proj{k}_{B}\cdot\tr\left[\left(\rho_{A}^{k}\right)^{0}\right]\right\|_{\infty}=\log\max_{k}\tr\left[\left(\rho_{A}^{k}\right)^{0}\right]=\max_{k}H_{0}(A)_{\rho^{k}}\ .
\end{align}
\end{proof}

Smooth entropy measures are defined by extremizing the non-smooth measures over a set of nearby states. Since we will later use some of the ideas from~\cite{datta-2008-7}, we use the same definitions as in~\cite{datta-2008-7}.

\begin{definition}
Let $\eps\geq0$, and $\rho_{AB}=\sum_{k}\rho_{A}^{k}\otimes\proj{k}_{B}\in\cS(\cH_{AB})$. The \textit{smooth alternative max-entropy} of $A$ conditioned on $B$ is defined as
\begin{align}
H_{0}^{\eps}(A|B)_{\rho}=\sup_{\overline{\rho}_{AB}\in\cB^{\eps}_{qc}(\rho_{AB})}H_{0}(A|B)_{\overline{\rho}}\ ,
\end{align}
where
\begin{align}
\cB^{\eps}_{qc}(\rho_{AB})=\{\bar{\rho}_{AB}\in\cP(\cH):\bar{\rho}_{AB}=\sum_{k}\bar{\rho}_{A}^{k}\otimes\proj{k}_{B},\|\rho_{AB}-\bar{\rho}_{AB}\|_{1}\leq\eps\}\ .
\end{align}
\end{definition}

In the technical part of this paper we will need distance measures. For $\rho,\sigma\in\cS_{\leq}(\cH)$ the purified distance is defined as~\cite[Definition 4]{Tomamichel09}
\begin{align}
P(\rho,\sigma)=\sqrt{1-\genFid^{2}(\rho,\sigma)}\ ,
\end{align}
where $\genFid(\cdot\,,\cdot)$ denotes the generalized fidelity (which equals the standard fidelity\footnote{The fidelity between $\rho,\sigma\in\cS_{\leq}(\cH)$ is defined as $F(\rho,\sigma)=\left\|\sqrt{\rho}\sqrt{\sigma}\right\|_1$.} if at least one of the states is normalized),
\begin{align}
\genFid(\rho,\sigma)=F(\rho,\sigma)+\sqrt{\left(1-\tr[\rho]\right)\left(1-\tr[\sigma]\right)}\ .
\end{align}
The purified distance is a metric on $\cS_{\leq}(\cH)$~\cite[Lemma 5]{Tomamichel09}. Henceforth we call $\rho$, $\sigma\in\cS_{\leq}(\cH)$ $\eps$-close if $P(\rho,\sigma)\leq\eps$ and denote this by $\rho\approx_{\eps}\sigma$. Furthermore, we will also need a distance measure for quantum channels. We use a norm on the set of CPTP maps which measures the probability by which two such mappings can be distinguished. The norm is known as the diamond norm in quantum information theory~\cite{Kitaev97}. Here, we present it in a formulation which highlights that it is dual to the well-known completely bounded (cb) norm~\cite{Paulsen}. 

\begin{definition}\label{def:diamond}
Let $\cE_{A}:\cL(\cH_{A})\mapsto\cL(\cH_{B})$ be a linear map. The diamond norm of $\cE_{A}$ is defined as
\begin{align}
\|\cE_{A}\|_{\diamond}=\sup_{k\in\mathbb{N}}\|\cE_{A}\otimes\cI_{k}\|_{1}\ ,
\end{align}
\end{definition}

The supremum in Definition~\ref{def:diamond} is reached for $k=|A|$~\cite{Kitaev97, Paulsen}. We call two CPTP maps $\cE$ and $\cF$ $\eps$-close if they are $\eps$-close in the metric induced by the diamond norm.

It is the main of result of~\cite{datta-2008-7} to quantify how much entanglement is needed in order to create a single copy of a bipartite state $\rho_{AB}$~\cite{HayashiBook}, a scenario previously studied in the asymptotic i.i.d.~setting~\cite{Hayden01,Bennett96}.

\begin{definition}\label{def:mainstate}
Consider a bipartite system with parties Alice and Bob, where Alice controls a system $\cH_{A}$ and Bob $\cH_{B}$. Let $\eps\geq0$, $\Phi_{\bar{A}\bar{B}}$ be a maximally entangled state between Alice and Bob, and $\rho_{AB}\in\cS(\cH_{AB})$. An $\eps$-faithful one-shot entanglement dilution protocol for $\rho_{AB}$ is a local operation and classical communication (LOCC) operation $\Lambda$ between Alice and Bob with $\bar{A}\rightarrow A$ at Alice's side and $\bar{B}\rightarrow B$ at Bob's side, such that
\begin{align}
\Lambda(\Phi_{\bar{A}\bar{B}})\approx_{\eps}\rho_{AB}\ .
\end{align}
If $\Phi_{\bar{A}\bar{B}}$ has Schmidt rank $R$, $\log R$ is the dilution cost of the one-shot entanglement dilution protocol.
\end{definition}

\begin{definition}\label{def:state_cost}
Let $\eps\geq0$ and $\rho_{AB}\in\cS(\cH_{AB})$. The minimal dilution cost of all $\eps$-faithful one-shot entanglement dilution protocols for $\rho_{AB}$ is called $\eps$-faithful \textit{one-shot entanglement cost} of $\rho_{AB}$ and is denoted by $E_{C}^{(1)}(\rho_{AB},\eps)$.
\end{definition}

\begin{proposition}\cite[Theorem 1]{datta-2008-7}\label{prop:datta}
Let $\eps\geq0$ and $\rho_{AB}\in\cS(\cH_{AB})$. Then,
\begin{align}\label{eq:mainstate}
\min_{\{p_{i},\rho^{i}\}}H_{0}^{2\sqrt{\eps}}(A|R)_{\rho}\leq E_{C}^{(1)}(\rho_{AB},\eps)\leq\min_{\{p_{i},\rho^{i}\}}H_{0}^{\eps/2}(A|R)_{\rho}\ ,
\end{align}
where the minimum ranges over all pure states decompositions $\rho_{AB}=\sum_{i}p_{i}\rho_{AB}^{i}$ and $\rho_{AR}=\sum_{i}p_{i}\rho^{i}_{A}\otimes\ket{i}\bra{i}_{R}$.
\label{nilanjana}
\end{proposition}

The idea for the achievability is as follows. For any pure state decomposition $\rho_{AB}=\sum_{i}p_{i}\rho_{AB}^{i}$ Alice can locally create the classical-quantum state $\rho_{ABR}=\sum_{i}p_{i}\rho_{AB}^{i}\otimes\proj{i}_{R}$, and then, conditioned on the index $i$, teleport the $B$-part of the pure states $\rho_{AB}^{i}$ to Bob. Minimizing over all pure state decompositions, a straightforward analysis shows that the resulting entanglement cost is bounded as in Proposition~\ref{prop:datta}. We will make use of these ideas for the proof of our main theorem.

\begin{remark}\label{rmk:datta}
The bounds given in~\eqref{eq:mainstate} also hold if we only allow one-way classical communication (forward or backward).
\end{remark}


\section{Entanglement Cost of Quantum Channels}\label{sec:main}

\subsection{Main Result}

We are now in the position to define the entanglement cost of quantum channels and prove the main result of this paper, Theorem \ref{main_1}.

\begin{definition}\label{def:simulation}
Consider a bipartite system with parties Alice and Bob. Let $\eps\geq0$, $\Phi_{\bar{A}\bar{B}}$ be a maximally entangled state between Alice and Bob, and $\cE:\cL(\cH_{A})\rightarrow\cL(\cH_{B})$ be a CPTP map, where Alice controls $\cH_{A}$ and Bob $\cH_{B}$. A one-shot channel simulation for $\cE$ with error $\eps$ is a quantum protocol
\begin{align}
\cF:\quad&\cL(\cH_{A})\rightarrow\cL(\cH_{B})\notag\\
&\rho_{A}\qquad\mapsto\Lambda(\rho_{A}\otimes\Phi_{\bar{A}\bar{B}})\ ,
\end{align}
where $\Lambda$ is a LOCC operation between Alice and Bob with $A\bar{A}\rightarrow0$ (no output) at Alice's side and $\bar{B}\rightarrow B$ at Bob's side, as well as
\begin{align}
\|\cF-\cE\|_{\Diamond}\leq\eps\ .
\end{align}
If $\Phi_{\bar{A}\bar{B}}$ has Schmidt rank $R$, $\log R$ is the entanglement cost of the one-shot channel simulation.
\end{definition}

By the definition of the diamond norm (Definition~\ref{def:diamond}), this assures that for any possible input state, the output of the channel simulation $\cF$ can only distinguished with small probability from the corresponding output of $\cE$. 

\begin{definition}\label{def:main}
Let $\cE:\cL(\cH_{A})\rightarrow\cL(\cH_{B})$ be a CPTP map. An asymptotic channel simulation for $\cE$ is a sequence of one-shot channel simulations $\cF^{n}$ for $\cE^{\otimes n}$ with error $\eps_{n}$, such that $\lim_{n\rightarrow\infty}\eps_{n}=0$. The entanglement cost of the simulation is $\limsup_{n\rightarrow\infty}\frac{\log R_{n}}{n}$.
\end{definition}

In the language of general channel simulations this corresponds to a so-called non-feedback simulation, since Alice does not obtain the output of the complementary channel~\cite{Bennett06}.

\begin{theorem}\label{main_1}
Let $\cE_{A\rightarrow B}:\cL(\cH_{A})\rightarrow\cL(\cH_{B})$ be a CPTP map. Then, the minimal entanglement cost $E_{C}(\cE_{A\rightarrow B})$ of an asymptotic channel simulation for $\cE_{A\rightarrow B}$ is given by
\begin{align}\label{eq:main}
E_{C}(\cE_{A\rightarrow B})=\lim_{n\rightarrow\infty}\frac{1}{n}\max_{\psi^{n}_{AA'}}E_{F}\left(\left(\cE^{\otimes n}_{A\rightarrow B}\otimes\cI_{A'}\right)\left(\psi^{n}_{AA'}\right)\right)\ ,
\end{align}
where $\psi^{n}_{AA'}=\cV(\cH_{A}^{\otimes n}\otimes\cH_{A'}^{\otimes n})$ and $\cH_{A'}\cong\cH_{A}$.
\end{theorem}

\begin{proof}
We first show that the right-hand side of~\eqref{eq:main} can be achieved (Proposition~\ref{final}), and thereafter that it is also a lower bound (Proposition~\ref{prop:converse}).
\end{proof}


\subsection{Proof: Achievability}\label{sec:achiev}

The proof proceeds in three steps leading to Proposition~\ref{final}. The basic idea is as follows. Given a quantum channel $\cE$, we need to show the existence of a sequence of one-shot channel simulations with asymptotically vanishing error, and an asymptotic entanglement cost upper bounded by the right-hand side of~\eqref{eq:main}. The crucial step is that by the post-selection technique for quantum channels (Proposition~\ref{posti}), it is sufficient to come up with CPTP map (which consists of using maximally entangled states, local operations, and classical communication) that works for the purification of one special de Finetti input state. For this we use ideas from the one-shot entanglement cost of quantum states (Proposition~\ref{prop:datta}).


\begin{lemma}\label{main_2}
Let $\cE_{A\rightarrow B}:\cL(\cH_{A})\rightarrow\cL(\cH_{B})$ be a CPTP map. Then,
\begin{align}\label{firststep}
E_{C}(\cE_{A\rightarrow B})\leq\inf_{\{M^{k}_{A\rightarrow B}\}}\sup_{\psi_{A}\in\cS(\cH_{A})}\sum_{k}p_{k}H(B)_{\psi^{k}}\ ,
\end{align}
where the infimum is over all Kraus decompositions $\{M^{k}_{A\rightarrow B}\}$ of $\cE_{A\rightarrow B}$, $\psi^{k}_{B}=\frac{1}{p_{k}}M^{k}_{A\rightarrow B}\psi_{A}\left.M^{k}_{A\rightarrow B}\right.^{\dagger}$ and $p_{k}=\tr\left[M^{k}_{A\rightarrow B}\psi_{A}\left.M^{k}_{A\rightarrow B}\right.^{\dagger}\right]$.
\end{lemma}

\begin{proof}
We construct a sequence of one-shot channel simulations $\cF^{n}$ with asymptotically vanishing error $\eps_{n}$, and an asymptotic entanglement cost $\frac{\log R_{n}}{n}$ as in~\eqref{firststep}. Without lost of generality we choose $\cF^{n}$ to be permutation-covariant.\footnote{This can be seen as follows. First, Alice and Bob create shared randomness using classical communication. Then, Alice applies a random permutation $\pi$ on the input system chosen according to the shared randomness. This is followed by the original map (which might not yet be permutation-covariant), and Bob who undoes the permutation by applying $\pi^{-1}$ on the output system. If needed, the classical communication cost of this procedure can be kept sub-linear in $n$ by using randomness recycling, as discussed in~\cite[Section IV. D]{Bennett06}. Alternatively, one could also use a sub-linear amount of entanglement to assure the permutation covariance.} The post-selection technique (Proposition~\ref{posti}) applies to permutation-covariant quantum channels and upper bounds the error by
\begin{align}\label{selection}
\eps_{n}=\left\|\cE_{A\rightarrow B}^{\otimes n}-\cF^{n}_{A\rightarrow B}\right\|_{\Diamond}\leq(n+1)^{|A|^{2}-1}\cdot\left\|\left(\left(\cE_{A\rightarrow B}^{\otimes n}-\cF^{n}_{A\rightarrow B}\right)\otimes\cI_{A'}^{\otimes n}\otimes\cI_{E}\right)\left(\zeta^{n}_{AA'E}\right)\right\|_{1}\ ,
\end{align}
where $\zeta^{n}_{AA'E}$ is a purification of the de Finetti state $\zeta^{n}_{AA'}=\int\psi_{AA'}^{\otimes n}d(\psi_{AA'})$ with $\psi_{AA'}\in\cV(\cH_{A}\otimes\cH_{A'})$, $\cH_{A'}\cong\cH_{A}$ and $d(\cdot)$ the measure on the normalized pure states on $\cH_{A}\otimes\cH_{A'}$ induced by the Haar measure on the unitary group acting on $\cH_{A}\otimes\cH_{A'}$, normalized to $\int d(\cdot)=1$. Hence it is sufficient that the channel simulation $\cF^{n}$ works on the state $\zeta^{n}_{AA'E}$ leading to
\begin{align}\label{structure}
\omega_{BA'E}^{n}=\left(\cE_{A\rightarrow B}^{\otimes n}\otimes\cI_{A'}^{\otimes n}\otimes\cI_{E}\right)\left(\zeta^{n}_{AA'E}\right)\ ,
\end{align}
up to an error $o\left((n+1)^{1-|A|^{2}}\right)$ in trace distance, for an asymptotic entanglement cost smaller than~\eqref{firststep}.\\

For $\{M_{A\rightarrow B}^{n,k}\}$ a Kraus decomposition of $\cE_{A\rightarrow B}^{\otimes n}$, Alice locally applies the CPTP map with Kraus operators $M_{A\rightarrow B}^{n,k}\otimes\ket{k}_R$ to the state $\zeta^{n}_{AA'E}$ and sends a copy of the classical register $k$ to Bob creating the state
\begin{align}
\omega_{A'BER}^n=\sum_{k}M_{A\rightarrow B}^{n,k}\zeta^{n}_{AA'E}\left.M_{A\rightarrow B}^{n,k}\right.^{\dagger}\otimes\proj{k}_{R}\ .
\end{align}
Conditioned on $k$ Alice and Bob can now use $\log \mathrm{rank}\left(M_{A\rightarrow B}^{n,k}\right)$ many ebits to teleport $B$ from Alice to Bob (since $\zeta^{n}_{AA'E}$ is pure). By a property of the alternative conditional max-entropy on quantum-classical states (Lemma~\ref{lem:h0class}) this then leads to a total entanglement cost of
\begin{align}
H_0(B|R)_{\omega^n}=\log\max_k\mathrm{rank}\left(M_{A\rightarrow B}^{n,k}\right)\ .
\end{align}
Moreover, at the cost of an approximation error $\delta_n\geq0$ in purified distance, this can be reduced to $H_0^{\delta_{n}/4}(B|R)_{\omega^n}$. This is achieved by pretending that we have another quantum-classical state $\bar{\omega}_{A'BER}^n$ which is $\delta_{n}/4$ close to $\omega_{A'BER}^n$ in trace distance, and then applying the teleportation protocol defined by $\bar{\omega}_{A'BER}^n$.\footnote{See Lemma~\ref{a:1} for the equivalence of distance measures.} Now by taking an infimum over all Kraus decomposition $\{M_{A\rightarrow B}^{n,k}\}$ of $\cE_{A\rightarrow B}^{\otimes n}$ we get a $\delta_{n}$-faithful (measured in purified distance) channel simulation of $\cF^{n}$ on the input state $\zeta^{n}_{AA'E}$ for an entanglement cost upper bounded by
\begin{align}\label{cruc}
\inf_{\{M_{A\rightarrow B}^{n,k}\}}H_{0}^{\delta_{n}/4}(B|R)_{\omega^{n}}\ .
\end{align}
Using a corollary of Carath\'{e}odory's theorem (Lemma~\ref{mario}), we know that 
\begin{align}
\zeta^{n}_{AA'}\equiv\int\psi_{AA'}^{\otimes n}d(\psi_{AA'})=\sum_{j=1}^{N}q_{j}\left(\psi_{AA'}^{j}\right)^{\otimes n}\ ,
\end{align}
with $\psi_{AA'}^{j}\in\cV(\cH_{A}\otimes\cH_{A'})$, $N=(n+1)^{2\left(|A|^{2}-1\right)}$ and $\{q_{j}\}_{j=1}^{N}$ a probability distribution. This allows us to write
\begin{align}
\omega_{BR}^{n}=\sum_{j=1}^{N}q_{j}\sum_{k}M_{A\rightarrow B}^{n,k}\left(\psi_{A}^{j}\right)^{\otimes n}\left.M_{A\rightarrow B}^{n,k}\right.^{\dagger}\otimes\proj{k}_{R}\ .
\end{align}
One particular choice for a Kraus decomposition $\{M_{A\rightarrow B}^{n,k}\}$ of $\cE_{A\rightarrow B}^{\otimes n}$ in~\eqref{cruc} is then to choose a Kraus decomposition $\{M_{A\rightarrow B}^{k}\}$ for $\cE_{A\rightarrow B}$ and take this decomposition for every tensor product factor. Thus we find a $\delta_{n}$-faithful (measured in purified distance) channel simulation of $\cF^{n}$ on the input state $\zeta^{n}_{AA'E}$ for an entanglement cost upper bounded by
\begin{align}
\inf_{\{M_{A\rightarrow B}^{k}\}}H_{0}^{\delta_{n}/4}(B|R)_{\omega^{n}}\ ,
\end{align}
where the infimum ranges over all Kraus decompositions $\{M_{A\rightarrow B}^{k}\}$ of $\cE_{A\rightarrow B}$, and $\omega_{BR}^{n}=\sum_{j=1}^{N}q_{j}\omega_{BR}^{j}$ with
\begin{align}
\omega^{j}_{BR}=\sum_{k}M_{A\rightarrow B}^{k}\psi_{A}^{j}\left.M_{A\rightarrow B}^{k}\right.^{\dagger}\otimes\proj{k}_{R}\ .
\end{align}
But by a property of the smooth alternative conditional max-entropy (Lemma~\ref{smoothi}) we have
\begin{align}
\inf_{\{M_{A\rightarrow B}^{k}\}}H_{0}^{\delta_{n}/4}(B|R)_{\omega^{n}}\leq\inf_{\{M_{A\rightarrow B}^{k}\}}\max_{j}H_{0}^{\delta_{n}/4}(B|R)_{\left(\omega^{j}\right)^{\otimes n}}+2\left(|A|^{2}-1\right)\cdot\log(n+1)\ .
\end{align}
Using the asymptotic equipartition property for the smooth alternative conditional max-entropy (Lemma~\ref{lem:aep}) we arrive at an entanglement cost of
\begin{align}\label{concluding}
n\cdot\left\{\inf_{\{M_{A\rightarrow B}^{k}\}}\max_{j}H(B|R)_{\omega^{j}}\right\}+\sqrt{n}\cdot\log\left(|B|+3\right)\cdot\sqrt{\log\left(\frac{16}{\delta_{n}^{2}}\right)}+2\left(|A|^{2}-1\right)\cdot\log(n+1)\ .
\end{align}
Now we choose $\delta_{n}=\frac{1}{2}(n+1)^{2\left(1-|A|^{2}\right)}$ and the entanglement cost becomes
\begin{align}
n\cdot\left\{\inf_{\{M_{A\rightarrow B}^{k}\}}\max_{j}H(B|R)_{\omega^{j}}\right\}+\sqrt{n}\cdot\log\left(|B|+3\right)\cdot\sqrt{2+4\cdot\log(n+1)\cdot\left(|A|^{2}-1\right)}+2\left(|A|^{2}-1\right)\cdot\log(n+1)\ .
\end{align}
By the equivalence of the purified distance and the trace distance (Lemma~\ref{a:1}), the error measured in the trace distance is then upper bounded by $(n+1)^{2\left(1-|A|^{2}\right)}$. This together with~\eqref{selection} implies that there exists a sequence of one-shot channel simulations $\cF^{n}$ for $\cE^{\otimes n}$ with error
\begin{align}
\lim_{n\rightarrow\infty}\eps_{n}=\lim_{n\rightarrow\infty}\|\cE_{A\rightarrow B}^{\otimes n}-\cF^{n}_{A\rightarrow B}\|_{\Diamond}\leq\lim_{n\rightarrow\infty}(n+1)^{1-|A|^{2}}=0\ ,
\end{align}
where the entanglement cost of this asymptotic channel simulation is bounded by
\begin{align}
\inf_{\{M_{A\rightarrow B}^{k}\}}\max_{j}H(B|R)_{\omega^{j}}\leq\inf_{\{M_{A\rightarrow B}^{k}\}}\sup_{\psi_{A}\in\cS(\cH_{A})}\sum_{k}p_{k}H(B)_{\psi^{k}}\ ,
\end{align}
where the infimum ranges over all Kraus decompositions $\{M_{A\rightarrow B}^{k}\}$ of $\cE_{A\rightarrow B}$, $\psi_{B}^{k}=\frac{1}{p_{k}}M_{A\rightarrow B}^{k}\psi_{A}\left.M_{A\rightarrow B}^{k}\right.^{\dagger}$ and $p_{k}=\tr\left[M_{A\rightarrow B}^{k}\psi_{A}\left.M_{A\rightarrow B}^{k}\right.^{\dagger}\right]$.



\end{proof}

\begin{lemma}\label{lem:nonblocking}
Let $\cE_{A\rightarrow B}:\cL(\cH_{A})\rightarrow\cL(\cH_{B})$ be a CPTP map. Then
\begin{align}
E_{C}(\cE_{A\rightarrow B})\leq E_{C}^{1}(\cE_{A\rightarrow B})\equiv\max_{\psi_{AA'}}E_{F}\left(\left(\cE_{A\rightarrow B}\otimes\cI_{A'}\right)\left(\psi_{AA'}\right)\right)\ ,
\end{align}
where $\psi_{AA'}\in\cV(\cH_{A}\otimes\cH_{A'})$ and $\cH_{A'}\cong\cH_{A}$.
\end{lemma}

\begin{proof}
The basic idea is to use a minimax theorem (Lemma~\ref{lem:minimax}) to interchange the infimum with the supremum in the preceding lemma (Lemma~\ref{main_2}). To start with, we want to discretize the set of Kraus decompositions $\{M_{k}\}$ of $\cE$ with at most $\chi$ Kraus operators. For this we note that every such Kraus decomposition $\{M_{k}\}$ can be seen as a vector $v_{\chi}\in\mathbb{C}^{\chi\cdot|A||B|}$, by just writing all Kraus operators one after another in a vector.\footnote{Kraus decompositions with less than $\chi$ Kraus operators can just be filled up with zeros.} Furthermore, we have $\sum_{k}M_{k}^{\dagger}M_{k}=\1_{B}$ and therefore $v_{\chi}\in\cN_{\chi}=\{w\in\mathbb{C}^{\chi\cdot|A||B|}\mid\|w\|_{2}=\sqrt{|B|}\}$.\footnote{For this note that $\|v_{\chi}\|_{2}=\|\sum_{k}M_{k}^{\dagger}M_{k}\|_{2}$, where the norm on the let hand side denotes the euclidean vector norm and the norm on the right hand side denotes the Hilbert-Schmidt matrix norm.} We now discretize the set $\cT_{\chi}\subseteq\cN_{\chi}$ of all $v_{\chi}$ that correspond to a Kraus decomposition $\{M_{k}\}$ of $\cE$ with at most $\chi$ Kraus operators, using a lemma about $\eps$-nets (Lemma~\ref{lem:net}). The lemma states that there exists a set $\cT_{\chi,\eps}\subseteq\cT_{\chi}$ with $|\cT_{\chi,\eps}|\leq\left(\frac{2\sqrt{|B|}}{\eps}+1\right)^{2\chi\cdot|A||B|}\equiv M(\chi,\eps)$, such that for every $v_{\chi}\in\cT_{\chi}$, there exists a $v_{\chi,\eps}\in\cT_{\chi,\eps}$ with $\|v_{\chi}-v_{\chi,\eps}\|_{2}\leq\eps$.\\

As the next step we consider the set $\Gamma_{\chi,\eps}$ of probability distributions $\{q_{j}\}_{j=1}^{N}$ over $\cT_{\chi,\eps}$, and note for every such probability distribution, there exists a corresponding Kraus decomposition $\{\sqrt{q_{j}}\cdot M_{j,k}\}_{j,k=1}^{N,\chi}$ of $\cE$. Restricting the infimum in~\eqref{firststep} to $\Gamma_{\chi,\eps}$, we find
\begin{align}\label{eq:secondstep}
E_{C}(\cE_{A\rightarrow B})\leq\inf_{\Gamma_{\chi,\eps}}\sup_{\psi}\sum_{j}q_{j}\sum_{k}p_{j,k}H(B)_{\psi^{j,k}}\ ,
\end{align}
where $\psi^{j,k}=\frac{1}{p_{j,k}}M_{j,k}\psi M_{j,k}^{\dagger}$, and $p_{j,k}=\tr\left[M_{j,k}\psi M_{j,k}^{\dagger}\right]$.\\

To apply the minimax theorem (Lemma~\ref{lem:minimax}) to interchange the infimum and the supremum in~\eqref{eq:secondstep}, we need to check all the conditions of Lemma~\ref{lem:minimax}.\\

$\cS(\cH_{A})$ is compact, convex set. To see that $\sum_{j}q_{j}\sum_{k}p_{j,k}H(B)_{\psi^{j,k}}$ is concave in $\psi_{A}$, we consider $\psi_{A}=r^{(1)}\psi_{A}^{1}+r^{(2)}\psi_{A}^{2}$ with $\psi_{A}^{1},\psi_{A}^{2}\in\cS_{=}(\cH_{A})$ and $r^{(1)}+r^{(2)}=1$. We define $\tilde{r}^{(1)}_{j,k}=\frac{r^{(1)}\cdot p^{(1)}_{j,k}}{p_{j,k}}$, $\tilde{r}^{(2)}_{j,k}=\frac{r^{(2)}\cdot p^{(2)}_{j,k}}{p_{j,k}}$ with $p^{(1)}_{j,k}=\tr\left[M_{j,k}\psi^{1}_{A}M_{j,k}^{\dagger}\right]$, $p^{(2)}_{j,k}=\tr\left[M_{j,k}\psi^{2}_{A}M_{j,k}^{\dagger}\right]$. Since $\tilde{r}^{(1)}_{j,k}+\tilde{r}^{(2)}_{j,k}=1$, we have by the concavity of the von Neumann entropy for $\psi_{B}^{1,j,k}=\frac{M_{j,k}\psi^{1}_{A}M_{j,k}^{\dagger}}{p^{(1)}_{j,k}}$, $\psi_{B}^{2,j,k}=\frac{M_{j,k}\psi^{2}_{A}M_{j,k}^{\dagger}}{p^{(2)}_{j,k}}$ that
\begin{align}
H(B)_{\psi^{j,k}}\geq\tilde{r}^{(1)}_{j,k}H(B)_{\psi^{1,j,k}}+\tilde{r}^{(2)}_{j,k}H(B)_{\psi^{2,j,k}}\ .
\end{align}
By multiplying this with $q_{j}\cdot p_{j,k}$ and taking the sum over all $j,k$ we conclude
\begin{align}
\sum_{j}q_{j}\sum_{k}p_{j,k}H(B)_{\psi^{j,k}}\geq r^{(1)}\cdot\sum_{j}q_{j}\sum_{k}p^{(1)}_{j,k}H(B)_{\psi^{1,j,k}}+r^{(2)}\cdot\sum_{j}q_{j}\sum_{k}p^{(2)}_{j,k}H(B)_{\psi^{2,j,k}}\ .
\end{align}
The function $\sum_{j}q_{j}\sum_{k}p_{j,k}H(B)_{\psi^{j,k}}$ is also continuous in $\psi_{A}$, since for any $\psi_{A}^{1},\psi_{A}^{2}\in\cS(\cH_{A})$ with $\|\psi^{1}_{A}-\psi^{2}_{A}\|_{1}\leq\delta$ for some $\delta>0$, it follows from the monotonicity of the trace norm under CPTP maps and the continuity of the conditional von Neumann entropy (Lemma~\ref{lem:fannes}) that
\begin{align}
|\sum_{j}q_{j}\sum_{k}p^{(1)}_{j,k}H(B)_{\psi^{1,j,k}}-\sum_{j}q_{j}\sum_{k}p^{(2)}_{j,k}H(B)_{\psi^{2,j,k}}|\leq4\delta\log|B|+2h(\delta)\ ,
\end{align}
where $h(\cdot)$ denotes the binary Shannon entropy.\\

$\Gamma_{\chi,\eps}$ is a compact, convex set. Moreover $\sum_{j}q_{j}\sum_{k}p_{j,k}H(B)_{\psi^{j,k}}$ is linear in $\{q_{j}\}$ and therefore in particular convex and continuous. By finally applying the minimax theorem (Lemma~\ref{lem:minimax}) in~\eqref{eq:secondstep}, we find
\begin{align}\label{eq:christandl1}
E_{C}(\cE_{A\rightarrow B})\leq\sup_{\psi}\inf_{\Gamma_{\chi,\eps}}\sum_{j}q_{j}\sum_{k}p_{j,k}H(B)_{\psi^{j,k}}\ .
\end{align}
Since the function is concave, the infimum is taken on an extreme point and hence
\begin{align}\label{eq:christandl2}
\inf_{\Gamma_{\chi,\eps}}\sum_{j}q_{j}\sum_{k}p_{j,k}H(B)_{\psi^{j,k}}=\inf_{\{M_{k}\}}\sum_{k}p_{k}H(B)_{\psi^{k}}\ ,
\end{align}
where the second infimum ranges over all Kraus decompositions $\{M_{k}\}\cong v_{\chi,\eps}\in\cT_{\chi,\eps}$ of $\cE$.\\

Now let $0<\eps\leq\frac{1}{2\chi|B|}$. As the next step we show that for every Kraus decomposition $\{M_{k}\}\cong v_{\chi}\in\cT$ of $\cE$, there exists a Kraus decomposition $\{M_{k,\eps}\}\cong v_{\chi,\eps}\in\cT_{\chi,\eps}$ of $\cE$, such that
\begin{align}\label{eq:finalprop}
|\sum_{k}p_{k,\eps}H(B)_{\psi^{k,\eps}}-\sum_{k}p_{k}H(B)_{\psi^{k}}|\leq8\eps\chi|B|\log|B|+2h(2\eps\chi|B|)\ ,
\end{align}
where $\psi^{k,\eps}=\frac{1}{p_{k,\eps}}M_{k,\eps}\psi M_{k,\eps}^{\dagger}$, $p_{k,\eps}=\tr\left[M_{k,\eps}\psi M_{k,\eps}^{\dagger}\right]$, and $h(\cdot)$ denotes the binary Shannon entropy. To see this, we rewrite~\eqref{eq:finalprop}, using Definition~\ref{def:formation}, to
\begin{align}\label{eq:ideas}
|\sum_{k}p_{k,\eps}H(B)_{\psi^{k,\eps}}-\sum_{k}p_{k}H(B)_{\psi^{k}}|=|H(B|R)_{\psi^{k,\eps}}-H(B|R)_{\psi^{k}}|\ ,
\end{align}
where $\psi^{k,\eps}_{BR}=\sum_{k}p_{k,\eps}\psi^{k,\eps}_{B}\otimes\proj{k}_{R}$ and $\psi^{k}_{BR}=\sum_{k}p_{k}\psi^{k}_{B}\otimes\proj{k}_{R}$. To estimate~\eqref{eq:ideas} we want to use the continuity of the conditional von Neumann entropy (Lemma~\ref{lem:fannes}), and for this we analyze
\begin{align}
\left\|\sum_{k}p_{k,\eps}\psi^{k,\eps}_{B}\otimes\proj{k}_{R}-\sum_{k}p_{k}\psi^{k}_{B}\otimes\proj{k}_{R}\right\|_{1}=\sum_{k}\left\|M_{k,\eps}\psi M_{k,\eps}^{\dagger}-M_{k}\psi M_{k}^{\dagger}\right\|_{1}\ .
\end{align}
By the triangle inequality for the trace norm, the equivalence of the trace norm and the Hilbert-Schmidt norm (Lemma~\ref{lem:12norm}), and the sub-multiplicativity of the Hilbert-Schmidt norm (Lemma~\ref{lem:sub}), we get
\begin{align}
\sum_{k}\left\|M_{k,\eps}\psi M_{k,\eps}^{\dagger}-M_{k}\psi M_{k}^{\dagger}\right\|_{1}&\leq\sum_{k}\left\|M_{k,\eps}\psi\left(M_{k,\eps}^{\dagger}-M_{k}^{\dagger}\right)\right\|_{1}+\left\|\left(M_{k,\eps}-M_{k}\right)\psi M_{k}^{\dagger}\right\|_{1}\\
&\leq\sqrt{|B|}\left(\sum_{k}\left\|M_{k,\eps}\psi\left(M_{k,\eps}^{\dagger}-M_{k}^{\dagger}\right)\right\|_{2}+\left\|\left(M_{k,\eps}-M_{k}\right)\psi M_{k}^{\dagger}\right\|_{2}\right)\\
&\leq\sqrt{|B|}\left(\sum_{k}\|M_{k,\eps}\|_{2}\cdot\|\psi\|_{2}\cdot\left\|M_{k,\eps}^{\dagger}-M_{k,\eps}^{\dagger}\right\|_{2}+\left\|M_{k,\eps}-M_{k}\right\|_{2}\cdot\|\psi\|_{2}\cdot\|M_{k,\eps}^{\dagger}\|_{2}\right)\\
&\leq\sqrt{|B|}\left(\eps\chi\sqrt{|B|}+\chi\sqrt{|B|}\eps\right)=2\eps\chi|B|\ .
\end{align}
Finally~\eqref{eq:finalprop} follows by the continuity of the conditional von Neumann entropy (Lemma~\ref{lem:fannes}). Thus we find together with~\eqref{eq:christandl1} and~\eqref{eq:christandl2} that
\begin{align}
E_{C}(\cE_{A\rightarrow B})\leq\sup_{\psi}\inf_{\{M_{k}\}}\sum_{k}p_{k}H(B)_{\psi^{k}}+8\eps\chi|B|\log|B|+2h(2\eps\chi|B|)\ ,
\end{align}
where the infimum goes over all Kraus decompositions $\{M_{k}\}\cong v_{\chi}\in\cT$ of $\cE$ and $h(\cdot)$ denotes the binary Shannon entropy. Finally note that
\begin{align}
\inf_{\{M_{k}\}}\sum_{k}p_{k}H(B)_{\psi^{k}}=E_{F}\left(\sum_{k}(M^{k}_{A\rightarrow B})\psi_{AA'}(M^{k}_{A\rightarrow B})^{\dagger}\right)\ ,
\end{align}
where the infimum ranges over all Kraus decompositions $\{M_{k}\}$ of $\cE$, $\psi_{AA'}\in\cV(\cH_{A}\otimes\cH_{A'})$, and $\cH_{A'}\cong\cH_{A}$. But this infimum is actually taken for a decomposition of size at most $|A|^{2}|B|^{2}$ (Lemma~\ref{lem:uhlman}). Thus, if we set $\chi=|A|^{2}|B|^{2}$ and let $\eps\rightarrow0$, we find
\begin{align}
E_{C}(\cE_{A\rightarrow B})\leq\sup_{\psi_{AA'}}E_{F}\left(\left(\cE_{A\rightarrow B}\otimes\cI_{A'}\right)\left(\psi_{AA'}\right)\right)\ ,
\end{align}
where $\psi_{AA'}\in\cV(\cH_{A}\otimes\cH_{A'})$ and $\cH_{A'}\cong\cH_{A}$. Since the entanglement of formation is continuous (Lemma~\ref{lem:nielsen}) and $\cS(\cH_{A})$ is compact, the supremum can be turned into a maximum.
\end{proof}

\begin{proposition}\label{final}
Let $\cE_{A\rightarrow B}:\cL(\cH_{A})\rightarrow\cL(\cH_{B})$ be a CPTP map. Then,
\begin{align}
E_{C}(\cE_{A\rightarrow B})\leq\lim_{n\rightarrow\infty}\frac{1}{n}\max_{\psi^{n}_{AA'}}E_{F}\left(\left(\cE^{\otimes n}_{A\rightarrow B}\otimes\cI_{A'}\right)\left(\psi^{n}_{AA'}\right)\right)\ ,
\label{form}
\end{align}
where $\psi^{n}_{AA'}\in\cV(\cH_{A}^{\otimes n}\otimes\cH_{A'}^{\otimes n})$ and $\cH_{A'}\cong\cH_{A}$.
\end{proposition}

\begin{proof}
This follows from standard blocking arguments as in~\cite{Schumacher98}. Namely, by applying the non-regularized achievability (Lemma~\ref{lem:nonblocking}) to the quantum channel $\cE_{A\rightarrow B}^{\otimes n}$ for some $n>1$, we get
\begin{align}
E_{C}(\cE_{A\rightarrow B}^{\otimes n})\leq\frac{1}{n}\max_{\psi_{AA'}^{n}}E_{F}\left(\left(\cE_{A\rightarrow B}^{\otimes n}\otimes\cI_{A'}\right)\left(\psi_{AA'}^{n}\right)\right)\ ,
\end{align}
where $\psi^{n}_{AA'}\in\cV(\cH_{A}^{\otimes n}\otimes\cH_{A'}^{\otimes n})$ and $\cH_{A'}\cong\cH_{A}$. Since $n\cdot E_{C}(\cE_{A\rightarrow B})\leq E_{C}(\cE_{A\rightarrow B}^{\otimes n})$,\footnote{This is immediate since a channel simulation for $\cE_{A\rightarrow B}^{\otimes n}$ is a channel simulation for $n$ copies of $\cE_{A\rightarrow B}$.} we get the claim by letting $n\rightarrow\infty$.
\end{proof}


\subsection{Proof: Converse}\label{sec:conv}

The idea of the proof of the converse is that any asymptotic channel simulation for $\cE_{A\rightarrow B}$ must be able to produce any states of the form $\left(\cE^{\otimes n}_{A\rightarrow B}\otimes\cI_{A'}^{\otimes n}\right)\left(\psi^{n}_{AA'}\right)$ for $n\rightarrow\infty$. But by the converse for the one-shot entanglement cost for quantum states (Proposition~\ref{nilanjana}) we have a lower bound on the entanglement that is needed to do this.

\begin{proposition}\label{prop:converse}
Let $\cE_{A\rightarrow B}:\cL(\cH_{A})\rightarrow\cL(\cH_{B})$ be a CPTP map. Then,
\begin{align}
E_{C}(\cE_{A\rightarrow B})\geq\lim_{n\rightarrow\infty}\frac{1}{n}\max_{\psi^{n}_{AA'}}E_{F}\left(\left(\cE^{\otimes n}_{A\rightarrow B}\otimes\cI_{A'}\right)\left(\psi^{n}_{AA'}\right)\right)\ .
\end{align}
where $\psi^{n}_{AA'}\in\cV(\cH_{A}^{\otimes n}\otimes\cH_{A'}^{\otimes n})$ and $\cH_{A'}\cong\cH_{A}$.
\end{proposition}

\begin{proof}
By the definition of an $\eps$-faithful one-shot channel simulation $\cF^{n}$ for $\cE^{\otimes n}$ (Definition~\ref{def:simulation}), we have that
\begin{align}
\left\|\cF^{n}-\cE^{\otimes n}\right\|_{\Diamond}\leq\eps\ .
\end{align}
This implies in particular that
\begin{align}
\max_{\psi^{n}_{AA'}}\left\|\left(\left(\cF_{A\rightarrow B}^{n}-\cE^{\otimes n}_{A\rightarrow B}\right)\otimes\cI_{A'}\right)\left(\psi^{n}_{AA'}\right)\right\|_{1}\leq\eps\ ,
\end{align}
where $\psi^{n}_{AA'}\in\cV(\cH_{A}^{\otimes n}\otimes\cH_{A'}^{\otimes n})$ and $\cH_{A'}\cong\cH_{A}$. Hence every $\eps$-faithful one-shot channel simulation $\cF^{n}$ for $\cE^{\otimes n}$ needs to be able to produce any state of the form $\left(\cE^{\otimes n}_{A\rightarrow B}\otimes\cI_{A'}\right)\left(\psi^{n}_{AA'}\right)$ up to an error $\eps$ (measured in trace distance). But by the definition of the one-shot entanglement cost for quantum states (Definition~\ref{def:state_cost}), the entanglement that is needed for this, is given by
\begin{align}
\max_{\psi^{n}_{AA'}}E_{C}^{(1)}\left(\left(\cE^{\otimes n}_{A\rightarrow B}\otimes\cI_{A'}\right)\left(\psi^{n}_{AA'}\right),\eps/2\right)\ ,
\end{align}
where $\psi^{n}_{AA'}\in\cV(\cH_{A}^{\otimes n}\otimes\cH_{A'}^{\otimes n})$ and $\cH_{A'}\cong\cH_{A}$.\footnote{The factor $1/2$ appears because the one shot entanglement cost for quantum states is defined in terms of the purified distance (Definition~\ref{def:state_cost}), cf.~Lemma~\ref{a:1} about the equivalence of distance measures.} Thus we find for the entanglement cost of asymptotic channel simulations for $\cE_{A\rightarrow B}$ that 
\begin{align}\label{eq:conv1}
E_{C}(\cE_{A\rightarrow B})\geq\lim_{\eps\rightarrow0}\lim_{n\rightarrow\infty}\frac{1}{n}\max_{\psi^{n}_{AA'}}E_{C}^{(1)}\left(\left(\cE_{A\rightarrow B}^{\otimes n}\otimes\cI_{A'}\right)\left(\psi^{n}_{AA'}\right),\eps/2\right)\ ,
\end{align}
where $\psi^{n}_{AA'}\in\cV_{\leq}(\cH_{A}^{\otimes n}\otimes\cH_{A'}^{\otimes n})$ and $\cH_{A'}\cong\cH_{A}$. But for $\omega_{BA'}^{n}=\left(\cE^{\otimes n}_{A\rightarrow B}\otimes\cI_{A'}\right)\left(\psi^{n}_{AA'}\right)$, the converse for the one-shot entanglement cost for quantum states (Proposition~\ref{nilanjana}) says that
\begin{align}\label{eq:conv2}
E_{C}^{(1)}(\omega_{BA'}^{n},\eps/2)\geq\min_{\{p_{i},\omega^{i}\}}H_{0}^{\sqrt{2\eps}}(B|R)_{\omega^{n}}\ ,
\end{align}
where the minimum ranges over all pure states decompositions $\omega_{BA'}^{n}=\sum_{i}p_{i}^{n}\omega^{n,i}_{BA'}$ and $\omega_{BR}^{n}=\sum_{i}p_{i}^{n}\omega^{n,i}_{B}\otimes\proj{i}_{R}$. Now let $\bar{\omega}_{BR}^{n}\in\cB_{qc}^{\sqrt{2\eps}}(\omega_{BR}^{n})$ such that $H_{0}^{\sqrt{2\eps}}(B|R)_{\omega^{n}}=H_{0}(B|R)_{\bar{\omega}^{n}}$. Because the alternative conditional max-entropy is lower bounded by the conditional von Neumann entropy (Lemma~\ref{max}), and since the conditional von Neumann entropy is continuous (Lemma~\ref{lem:fannes}), we find
\begin{align}
H_{0}^{\sqrt{2\eps}}(B|R)_{\omega^{n}}=H_{0}(B|R)_{\bar{\omega}^{n}}\geq H(B|R)_{\bar{\omega}^{n}}\geq H(B|R)_{\omega^{n}}-4n\sqrt{2\eps}\log|B|-2h(\sqrt{2\eps})\ ,
\end{align}
where $h(\cdot)$ denotes the binary Shannon entropy. Thus, we conclude by the definition of the entanglement of formation (Definition~\ref{def:formation}) that
\begin{align}
\min_{\{p_{i}^{n},\omega^{n,i}\}}H_{0}^{\sqrt{2\eps}}(B|R)_{\omega^{n}}&\geq\min_{\{p_{i}^{n},\omega^{n,i}\}}H(B|R)_{\omega^{n}}-4n\sqrt{2\eps}\log|B|-2h(\sqrt{2\eps})\\
&=E_{F}(\omega_{BA'}^{n})-4n\sqrt{\eps}\log|B|-2h(\sqrt{2\eps})\\
&=E_{F}\left(\left(\cE^{\otimes n}_{A\rightarrow B}\otimes\cI_{A'}\right)\left(\psi^{n}_{AA'}\right)\right)-4n\sqrt{2\eps}\log|B|-2h(\sqrt{2\eps})\ ,
\end{align}
where the minimum ranges over all pure states decompositions $\omega_{BA'}^{n}=\sum_{i}p_{i}^{n}\omega^{n,i}_{BA'}$ and $\omega_{BR}^{n}=\sum_{i}p_{i}^{n}\omega^{n,i}_{B}\otimes\proj{i}_{R}$, as well as $\psi^{n}_{AA'}\in\cV(\cH_{A}^{\otimes n}\otimes\cH_{A'}^{\otimes n})$ with $\cH_{A'}\cong\cH_{A}$. Together with~\eqref{eq:conv1} and~\eqref{eq:conv2} this then implies
\begin{align}
E_{C}(\cE_{A\rightarrow B})&\geq\lim_{\eps\rightarrow0}\lim_{n\rightarrow\infty}\left\{\frac{1}{n}\max_{\psi^{n}_{AA'}}E_{F}\left(\left(\cE^{\otimes n}_{A\rightarrow B}\otimes\cI_{A'}\right)\left(\psi^{n}_{AA'}\right)\right)-4\sqrt{2\eps}\log|B|-\frac{2}{n}h(\sqrt{2\eps})\right\}\\
&=\lim_{n\rightarrow\infty}\frac{1}{n}\max_{\psi^{n}_{AA'}}E_{F}\left(\left(\cE^{\otimes n}_{A\rightarrow B}\otimes\cI_{A'}\right)\left(\psi^{n}_{AA'}\right)\right)\ ,
\end{align}
where $\psi^{n}_{AA'}\in\cV(\cH_{A}^{\otimes n}\otimes\cH_{A'}^{\otimes n})$ and $\cH_{A'}\cong\cH_{A}$. 
\end{proof}


\subsection{Properties}

Our main result (Theorem~\ref{main_1}) remains true if we restrict the classical communication to be one-way (forward or backward). This follows from the corresponding result about the entanglement cost of quantum states (Remark~\ref{rmk:datta}).\footnote{This is also true we think of the problem as simulating a noisy quantum channel from a perfect quantum channel (instead of simulating a noisy quantum channel from perfect entanglement), since in this case a maximally entangled state can always be distributed by the ideal channel.} We also note that the non-regularized achievability (Lemma~\ref{lem:nonblocking}) together with the converse (Proposition~\ref{prop:converse}) imply the following bounds.

\begin{corollary}\label{cor:non}
Let $\cE_{A\rightarrow B}:\cL(\cH_{A})\rightarrow\cL(\cH_{B})$ be a CPTP map. Then, we have that
\begin{align}\label{cor:prop}
\max_{\psi_{AA'}}E_{C}\left(\left(\cE_{A\rightarrow B}\otimes\cI_{A'}\right)\left(\psi_{AA'}\right)\right)\leq E_{C}(\cE_{A\rightarrow B})\leq\max_{\psi_{AA'}}E_{F}\left(\left(\cE_{A\rightarrow B}\otimes\cI_{A'}\right)\left(\psi_{AA'}\right)\right)\ ,
\end{align}
where $\psi_{AA'}\in\cV(\cH_{A}\otimes\cH_{A'})$, and $\cH_{A'}\cong\cH_{A}$.
\end{corollary}

Since the right-hand side of~\eqref{cor:prop} vanishes for every entanglement breaking channel,\footnote{A quantum channel $\cE_{A\rightarrow B}$ is called entanglement breaking if $\left(\cE_{A\rightarrow B}\otimes\cI_{A'}\right)\left(\psi_{AA'}\right)$ is separable for all $\psi_{AA'}\in\cV(\cH_{A}\otimes\cH_{A'})$.} and since the left-hand side of~\eqref{cor:prop} is greater than zero if the channel is not entanglement breaking~\cite{yang05}, this results in the following corollary.

\begin{corollary}
Let $\cE_{A\rightarrow B}:\cL(\cH_{A})\rightarrow\cL(\cH_{B})$ be a CPTP map. Then $E_{C}(\cE_{A\rightarrow B})=0$ if and only if $\cE_{A\rightarrow B}$ is entanglement breaking.
\end{corollary}


\section{Applications and Examples}\label{sec:app}

In this section we present two applications of our formula for the entanglement cost of channels and calculate some examples. We start with problem of proving security in the noisy storage model and then turn to the problem of deriving bounds for the strong converse of quantum capacities.


\subsection{Security in the Noisy Storage Model}\label{sec:noisy}

We will see below that $E_C$ forms a natural quantity when considering security in the noisy-storage model~\cite{Wullschleger09,Noisy1,steph:diss}. It will enable us to extend the parameter regime where security of all existing protocols~\cite{Wullschleger09,Curty10,noisy:robust,Noisy1,serge:new,serge:bounded,chris:tweak} can be proven. The appeal of this model is that it allows to solve any cryptographic problem involving two mutually distrustful parties, such as bit commitment, oblivious transfer~\cite{Wullschleger09} or secure identification~\cite{chris:id1,chris:id2}. This is impossible without imposing any assumptions, such as a noisy quantum memory, on the adversary~\cite{lo:insecurity,mayers:trouble,lo&chau:bitcom2,lo&chau:bitcom,mayers:bitcom}. Proposed protocols can thereby be implemented with any hardware suitable for quantum key distribution.\\

\begin{center}
\begin{figure}[ht]
\includegraphics[scale=0.7]{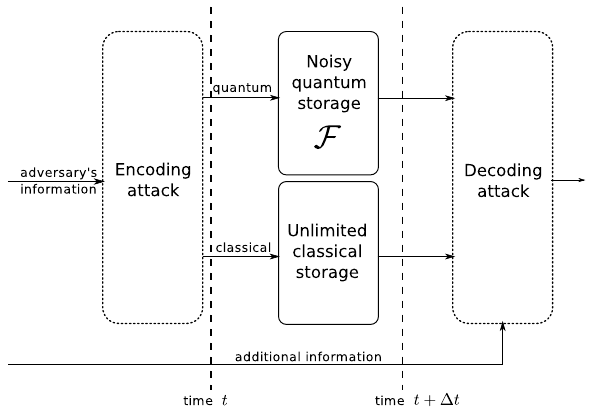}
\caption{Noisy-storage assumption: During waiting times $\Delta t$, the adversary can only use his noisy memory device to store quantum information. However, he is otherwise all powerful, and storage of classical information is free.}\label{fig:noisyModel}
\end{figure}
\end{center}

Let us first provide a brief overview of the noisy-storage model as illustrated in Figure~\ref{fig:noisyModel} - details can be found in e.g.~\cite{Wullschleger09}. The central assumption of the noisy-storage model is that the adversary can only store quantum information in a memory described by a particular channel $\cF:\cL(\cH_{in}) \rightarrow \cL(\cH_{out})$. In practice, the use of the memory device is enforced by introducing waiting times $\Delta t$ into the protocol. This is the only restriction imposed on the adversary who is otherwise all-powerful. In particular, he can store an unlimited amount of classical information, and all his actions are instantaneous. This includes any computations, communications, measurements and state preparation that may be necessary to perform an error-correcting encoding and decoding before and after using his noisy memory device.

In~\cite{Wullschleger09}, a natural link was formed between security in the noisy-storage model, and the information carrying capacity of the storage channel $\cF$. Of particular interest were thereby memory assumptions that scale with the number $m$ of qubits transmitted during the protocol.\footnote{In turn, this tells us how many qubits need to be send in order to achieve security against an attacker with a certain amount of storage.} That is, the channel is of the form $\cF = \cE^{\otimes \nu \cdot m}$, where $\nu$ is referred to as the storage rate. It was shown that any two-party cryptographic problem can in principle\footnote{That is, by transmitting a sufficiently large number $m$ of qubits.} be implemented securely if~\cite{Wullschleger09}
\begin{align}
C({\cE}) \cdot \nu < \frac{1}{2}\ ,
\end{align}
where $C(\cE)$ denotes the strong converse classical capacity of the channel $\cE$ (which is known to equal the classical capacity for certain classes of channels~\cite{Wehner_K09}). For the special case of $\cE = \cI_2$, i.e.~the one qubit identity channel, the condition simplifies to
\begin{align}\label{eq:BoundedSec}
\nu < \frac{1}{2}\ .
\end{align}
This case is also known as bounded-storage~\cite{serge:bounded,serge:new,chris:diss}. For protocols involving qubits in a simple BB84 like scheme this is the best bound known today, although using a protocol with very high dimensional encodings can lead to an improvement up to $\nu < 1$~\cite{prabha:limits}.\\

When considering storing quantum information exchanged during the protocol, it may come as a surprise that the classical capacity should be relevant. Indeed, looking at Figure~\ref{fig:noisyModel} it becomes clear that a much more natural quantity would be the quantum capacity of $\cE$. Whereas we do not accomplish this goal, we make significant progress by linking the security to $E_C(\cE)$.

\begin{lemma}
Let $m$ be the number of qubits transmitted in the protocol, and let the adversary's storage be of the form $\cF = \cE^{\otimes \nu\cdot m}$. Then for sufficiently large $m$ any two-party cryptographic primitive can be implemented securely in the noisy-storage model if
\begin{align}
E_C(\cE) \cdot \nu < \frac{1}{2}\ .
\end{align}
\end{lemma}

\begin{proof}
Consider the case of bounded, noise-free, memory. Note that~\eqref{eq:BoundedSec} from~\cite{Wullschleger09} tells us that security can be achieved for large enough $m$ if the dimension $d$ of the adversary's storage device is strictly smaller than $d < 2^{m/2}$. Now, suppose by contradiction that security could not be achieved with a storage of the form $\cF = \cE^{\otimes n}$, where $n=\nu\cdot m$ and $E_C(\cE) \cdot n \leq \log d$. However, then there exists a successful cheating strategy also in the case of bounded storage of dimension $d$: the adversary could simply simulate $\cE^{\otimes n}$ using an entangled state of dimension $d$ with $\log d = E_C(\cE)\cdot n$, possibly using additional classical forward communication provided by his unlimited classical storage device. Hence for large enough $m$, security can be achieved if $E_C(\cE) \cdot \nu < \frac{1}{2}$ as claimed.
\end{proof}

Note that for small $m$, a corresponding one-shot quantity $E_{C}^{(1)}$ is relevant (but is not discussed in this work).\footnote{However, statements for any finite $m$ can be made using our results (although the resulting bounds might not be optimal).} It should also be noted that our bound provides a further improvement apart from replacing $C$ by $E_{C}$, as we no longer explicitly require any strong converse behavior. This is implicitly provided by our simulation argument.\\

\begin{figure}[ht]
\begin{minipage}[b]{0.28\linewidth}
\centering
\includegraphics[scale=0.4]{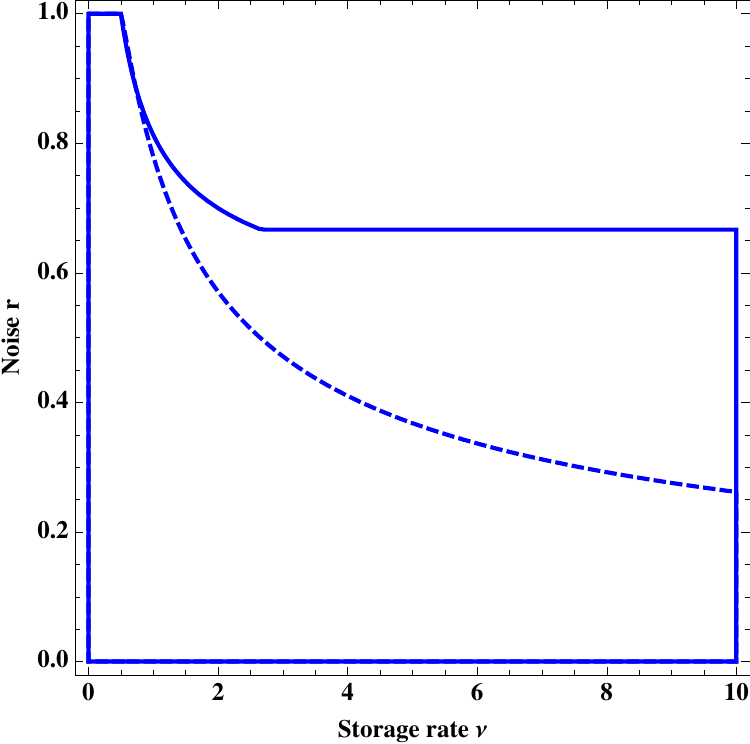}
\caption{Depolarizing channel. Security was previously known below the dashed line. Now for $(r,\nu)$ inside the solid line.}
\label{fig:depolImprove}
\end{minipage}
\hspace{0.2cm}
\begin{minipage}[b]{0.28\linewidth}
\centering
\includegraphics[scale=0.4]{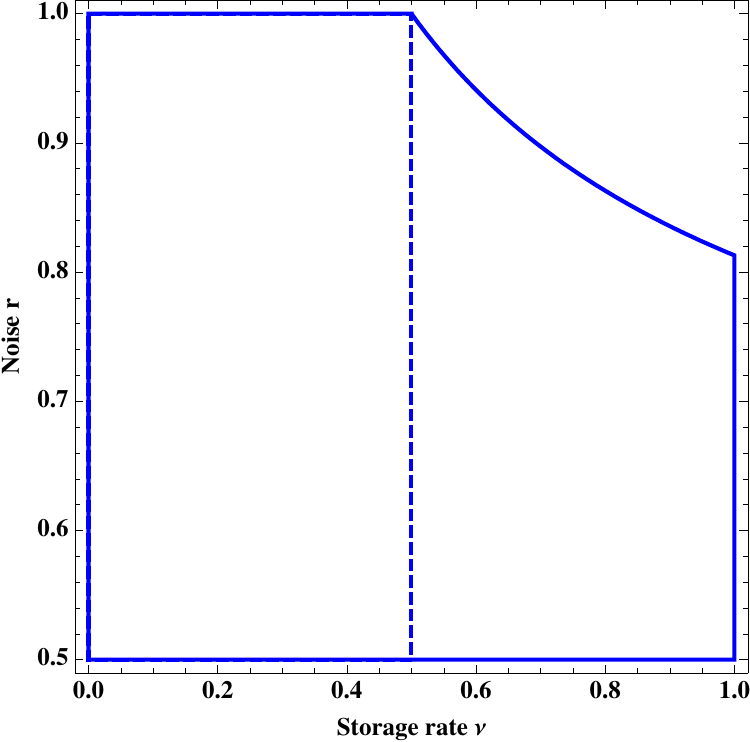}
\caption{Dephasing channel. Before security was no better than for bounded storage, left of dashed line. Now for $(r,\nu)$ inside the solid line.}
\label{fig:dephaseImprove}
\end{minipage}
\hspace{0.2cm}
\begin{minipage}[b]{0.28\linewidth}
\centering
\includegraphics[scale=0.4]{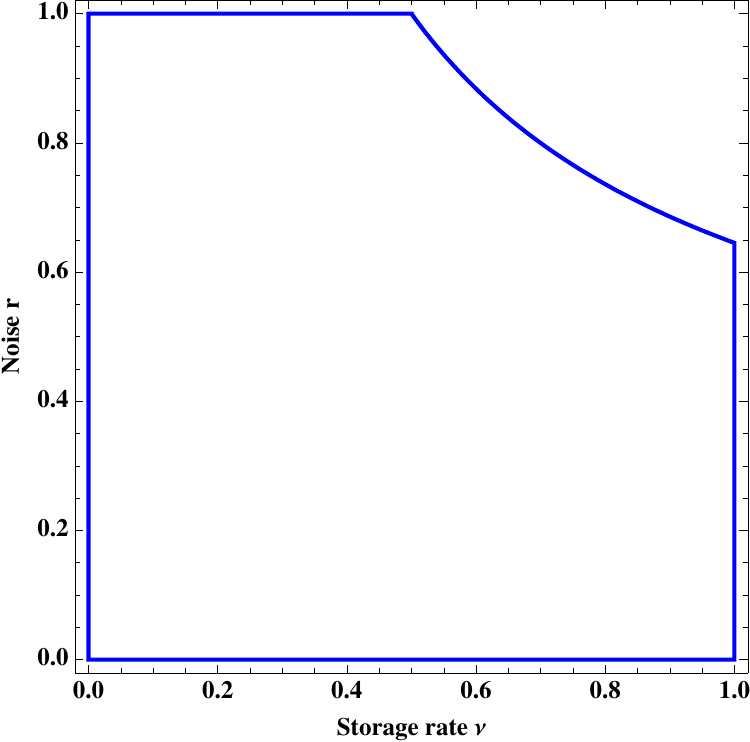}
\caption{Amplitude damping channel. No security statement was known previously. Now for $(r,\nu)$ inside the solid line.}
\label{fig:damping}
\end{minipage}
\end{figure}

At first glance, our improved bound may appear rather unsatisfying. How could we hope to use this bound to make explicit statements when the formula for $E_{C}$ involves regularization? First of all, note that for any entanglement breaking channel $\cE$, $E_{C}(\cE) = 0$, which leads to immediate security bounds: security can then be attained for any storage rate $\nu$. However, we can show security even for a much larger class of entanglement preserving channels. We now show that even though it is unclear how to calculate $E_{C}$ explicitly, we can nevertheless obtain improved bounds. The key to such bounds is Lemma~\ref{lem:nonblocking}, which gives us
\begin{align}
E_{C}(\cE_{A\rightarrow B})\leq E_{C}^{1}(\cE_{A\rightarrow B})=\max_{\psi_{AA'}}E_{F}\left(\left(\cE_{A\rightarrow B}\otimes\cI_{A'}\right)\left(\psi_{AA'}\right)\right)\ ,
\end{align}
where $\psi_{AA'}\in\cV(\cH_{A}\otimes\cH_{A'})$ and $\cH_{A'}\cong\cH_{A}$. Most channels considered in the noisy-storage model are qubit channels, and for these an exact formula for the entanglement of formation was shown in~\cite{Wootters98}
\begin{align}\label{eq:wootters}
E_{F}\left(\left(\cE_{A\rightarrow B}\otimes\cI_{A'}\right)\left(\psi_{AA'}\right)\right)=h\left(\frac{1}{2}+\frac{1}{2}\cdot\sqrt{1-C^{2}\left(\left(\cE_{A\rightarrow B}\otimes\cI_{A'}\right)\left(\psi_{AA'}\right)\right)}\right)\ ,
\end{align}
with $h(\cdot)$ the binary Shannon entropy, and the concurrence
\begin{align}\label{eq:Cexpr}
C(\rho)=\max\left\{0,\sqrt{\lambda_{1}}-\sqrt{\lambda_{2}}-\sqrt{\lambda_{3}}-\sqrt{\lambda_{4}}\right\}\ ,
\end{align}
with $\lambda_{i}$'s the eigenvalues of $\rho\tilde{\rho}$ in decreasing order, $\tilde{\rho}=(\sigma_{y}\otimes\sigma_{y})\rho^{*}(\sigma_{y}\otimes\sigma_{y})$ with $\rho^{*}$ the complex conjugate of $\rho$ in the canonical basis, and $\sigma_{y}=\begin{pmatrix}0 &-i\\i&0\end{pmatrix}$. Furthermore we know from~\cite{Verstraete01,Konrad08} that for $\psi_{AA'}$ pure
\begin{align}
C\left(\left(\cE_{A\rightarrow B}\otimes\cI_{A'}\right)\left(\psi_{AA'}\right)\right)=C\left(\left(\cE_{A\rightarrow B}\otimes\cI_{A'}\right)\left(\phi_{AA'}\right)\right)\cdot C(\psi_{AA'})\ ,
\end{align}
where $\phi_{AA'}$ denotes the maximally entangled state. Since $C(\psi_{AA'})\leq1$, it follows
\begin{align}\label{eq:secLimits}
E_{C}^{1}(\cE_{A\rightarrow B})=h\left(\frac{1}{2}+\frac{1}{2}\cdot\sqrt{1-C^{2}\left(\left(\cE_{A\rightarrow B}\otimes\cI_{A'}\right)\left(\phi_{AA'}\right)\right)}\right)\ ,
\end{align}
that is, it only remains to compute $C(\cdot)$ for the Choi-Jamiolkowski state of the channel. This can be done explicitly using~\eqref{eq:Cexpr} for any qubit channel of interest. To obtain a bound for when security can be achieved we thus can calculate when the condition
\begin{align}
	\nu \cdot h\left(\frac{1}{2}+\frac{1}{2}\cdot\sqrt{1-C^{2}\left(\left(\cE_{A\rightarrow B}\otimes\cI_{A'}\right)\left(\phi_{AA'}\right)\right)}\right) < \frac{1}{2}
\end{align}
is fulfilled. Figures~\ref{fig:depolImprove} and~\ref{fig:dephaseImprove} illustrate the improvements obtained for depolarizing and dephasing noise respectively. Note that since previous bounds involved the classical capacity, dephasing noise was no better than mere bounded storage. Using our new bound, however, we obtain non-trivial bounds even for this case. Figure~\ref{fig:damping} provides security bounds for the one qubit amplitude damping channel $\cE_{\rm damp}(\rho) = E_0 \rho E_0 + E_1 \rho E_1$ where $E_0 = \begin{pmatrix}  1&0\\0&\sqrt{r}\end{pmatrix}$ and $E_1 = \begin{pmatrix}0 & \sqrt{1-r}\\0&0\end{pmatrix}$. No previous security bound was known for this channel.

\subsection{An Upper Bound on the Strong Converse Quantum Capacity}\label{sec:strongconv}

To determine a quantum channel's capacity for sending information, two aspects need to be addressed. First of all, one needs to show that the capacity can be achieved. That is, there exists some coding scheme that allows to transmit information reliably at any rate up to the capacity. Second, however, the capacity should really form a threshold for information transmission. That is, if one tries to send information at a rate above the capacity, then there exists no coding scheme that allows to send information without any error. Such a statement is also known as a weak converse.\\

This however, does not yet exclude the possibility of sending information with a small error at a rate that exceeds the capacity. The minimal rate for which the success in transmitting information drops exponentially with the number of channel uses, is known as the strong converse capacity. The strong converse capacity is appealing since it really gives a sharp threshold for information transmission. But to determine the strong converse capacity forms a challenge even when it comes to sending classical information. Only when restricted to non-entangled input states~\cite{winter99,ogawa99} or certain classes of quantum channels~\cite{Wehner_K09}, it is known that the strong converse classical capacity is actually the same as the classical capacity. However, various upper bounds on the strong converse classical capacity are known~\cite{Bennett06,fernando1,ciara:converse}. For example, the quantum reverse Shannon theorem shows that the entanglement assisted classical capacity $C_{E}$ and its strong converse version are identical~\cite{Bennett06}. Of course $C_{E}$ is then also an upper bound on the unassisted strong converse classical capacity. In addition, the result immediately implies that the entanglement assisted quantum capacity $Q_{E}=C_{E}/2$ and its strong converse version are identical. Thus, $Q_{E}$ is an upper bound on the unassisted strong converse quantum capacity. \\

As the second application of our result, we prove a new upper bound on the strong converse quantum capacity. Similar to the quantum reverse Shannon theorem~\cite{Bennett06}, we employ the idea of a channel simulation to prove that when we send quantum information at a rate exceeding $E_{C}$, then the fidelity gets exponentially small. Our bound holds for all channels. To start with, let us first define the notion of quantum capacity more formally.

\begin{definition}
Consider a bipartite system with parties Alice and Bob. Let $\eps\geq0$ and $\cE:\cL(\cH_{A})\rightarrow\cL(\cH_{B})$ be a CPTP map, where Alice controls $\cH_{A}$ and Bob $\cH_{B}$. An $\eps$-error code for $\cE$ consists of an encoding CPTP map $\Lambda_{\mathrm{enc}}:\left(\mathbb{C}^{2}\right)^{\otimes R}\rightarrow\cH_{A}$ on Alice's side, and a decoding CPTP map $\Lambda_{\mathrm{dec}}:\cH_{B}\rightarrow\left(\mathbb{C}^{2}\right)^{\otimes R}$ on Bob's side such that
\begin{align}\label{eq:capacity}
\|\Lambda_{\mathrm{dec}}\circ\cE\circ\Lambda_{\mathrm{enc}}-\cI\|_{\Diamond}\leq\eps\ ,
\end{align}
where $\cI:\left(\mathbb{C}^{2}\right)^{\otimes R}\rightarrow\left(\mathbb{C}^{2}\right)^{\otimes R}$ is the identity channel, and the rate of the code is given by $R$. Furthermore, an asymptotic code for $\cE$ is a sequence of $\eps_{n}$-error codes for $\cE^{\otimes n}$ with rate $R_{n}$ such that $\lim_{n\rightarrow\infty}\eps_{n}=0$, and the corresponding asymptotic rate is given by $R=\limsup_{n\rightarrow\infty}\frac{R_{n}}{n}$. The quantum capacity $Q(\cE)$ is then defined as the minimal asymptotic rate of asymptotic codes for $\cE$.
\end{definition}

Note that there are slightly different ways to define the quantum capacity, and we could use other distance measures (like the entanglement fidelity or the channel fidelity) in~\eqref{eq:capacity}. Yet, it was as shown that all definitions lead to the same capacity  (see Lemma~\ref{lem:werner}, taken from~\cite{werner04}). Similarly, we can define the quantum capacity in the presence of free classical forward communication from the sender to the receiver, denoted by $Q_{\rightarrow}$, the quantum capacity in the presence of free classical backward communication from the receiver to the sender,  denoted by $Q_{\leftarrow}$, and the two-way classical communication assisted quantum capacity $Q_{\leftrightarrow}$.\\

As our argument makes crucial use of the idea of simulating a noisy channel with perfect, noise-free, channels, we now first establish a strong converse for the identity channel. For the unassisted quantum capacity this is straightforward, and can be understood in terms of the impossibility of compressing $n$ qubits into a smaller storage device.

\begin{lemma}\label{lem:identity}
Let $\cI_{2}$ be the qubit identity channel. Then we have for every sequence of $\eps_{n}$-error codes for $\cI_{2}^{\otimes n}$ with asymptotic rate $R$ that
\begin{align}
\eps_{n}\geq1-2^{-n(R-1)}\ .
\end{align}
\end{lemma}

\begin{proof}
For Kraus decompositions $\{E_{j}\}$, $\{D_{k}\}$ of the CPTP maps $\Lambda_{\mathrm{enc}}$, $\Lambda_{\mathrm{dec}}$ respectively, we get for the channel fidelity
\begin{align}
F_{c}(\Lambda_{\mathrm{dec}}\circ\cI\circ\Lambda_{\mathrm{enc}})&=\sum_{j,k}\left|\tr\left[D_{k}E_{j}\left(\frac{\1}{2^{nR}}\right)\right]\right|^{2}\leq\sum_{j,k}\tr\left[D_{k}E_{j}\left(\frac{\1}{2^{nR}}\right)E_{j}^{\dagger}D_{k}^{\dagger}\right]\tr\left[\Pi_{k}\left(\frac{\1}{2^{nR}}\right)\right]\\
&\leq\frac{1}{2^{nR}}\sum_{j,k}\tr\left[D_{k}E_{j}\left(\frac{\1}{2^{nR}}\right)E_{j}^{\dagger}D_{k}^{\dagger}\right]\tr\left[\Pi_{k}\right]\leq2^{-n(R-1)}\ ,
\end{align}
where $\Pi_{k}$ denotes the projector onto the subspace to which $D_{k}$ maps, and the first inequality follows from the Cauchy-Schwarz inequality. By $F_{c}(\cE)\geq1-\|\cE-\cI\|_{\diamond}$ (Lemma~\ref{lem:werner}) this implies the claim.
\end{proof}

This can be generalized to the case of free classical communication assistance.

\begin{corollary}\label{cor:strong}
Let $\cI_{2}$ be the qubit identity channel. Then we have for every sequence of classical communication assisted $\eps_{n}$-error codes for $\cI_{2}^{\otimes n}$ with asymptotic rate $R$ that
\begin{align}
\eps_{n}\geq1-2^{-n(R-1)}\ .
\end{align}
\end{corollary}

\begin{proof}
Since back communication is allowed, the general form of a protocol consists of potentially many rounds of forward quantum and classical communication as well as backward classical communication. We first analyze one such round, which has without lost of generality the following form:
\begin{enumerate}
\item CPTP map $\cD^{1}$ at the receiver with Kraus operators $\{D_{i}^{1}\}$
\item Classical communication from the receiver to the sender, denoted by the register $B$
\item CPTP map $\cE$ at the sender with Kraus operators $\{\hat{E}_{j,b}\}=\{E_{j,b}\otimes\proj{b}_{B}\}$
\item Classical communication from the sender to the receiver, denoted by the register $F$
\item CPTP map $\cD^{2}$ at the receiver with Kraus operators $\{\hat{D}_{k,f}^{2}\}=\{D_{k,f}^{2}\otimes\proj{f}_{F}\}$
\end{enumerate}
The channel fidelity after this round can be estimated as before (Lemma~\ref{lem:identity})
\begin{align}
F_{c}(\cD^{2}\circ(\cI_{2}^{\otimes n}\otimes\cI_{F})\circ\cE\circ\cI_{B}\circ\cD^{1})&=\sum_{ijkbf}\left|\tr\left[\hat{D}^{2}_{k,f}\hat{E}_{j,b}D_{i}^{1}\left(\frac{\1}{2^{nR}}\right)\right]\right|^{2}\\
&\leq\sum_{ijkbf}\tr\left[\hat{D}^{2}_{k,f}\hat{E}_{j,b}D_{i}^{1}\left(\frac{\1}{2^{nR}}\right)\left(D_{i}^{1}\right)^{\dagger}\hat{E}_{j,b}^{\dagger}\left(\hat{D}_{k,f}^{2}\right)^{\dagger}\right]\tr\left[\Pi_{k,f}\left(\frac{\1}{2^{nR}}\right)\right]\\
&\leq2^{-n(R-1)}\ ,
\end{align}
where $\Pi_{k,f}$ denote the projector onto the subspace that $\hat{D}^{2}_{k,f}$ maps. It is now easily seen that adding more rounds does not affect the argument; the projectors $\Pi$ are just chosen such that they project on the subspaces to which the Kraus operators of the last CPTP map at the receiver map to.
\end{proof}

To generalize this to arbitrary quantum channels we need one more ingredient. We need to show that the asymptotic channel simulation for some quantum channel (as discussed in Theorem~\ref{main_1}) can be done for an error rate which is exponentially small in $n$.

\begin{lemma}\label{prop:error}
Let $\cE:\cL(\cH_{A})\rightarrow\cL(\cH_{B})$ be a CPTP map and $\delta_{1}>0$. Then, there exists an asymptotic channel simulation for $\cE$ with an entanglement cost of $E_{C}+\delta_{1}$ and an error
\begin{align}
\alpha_{n}=(n+1)^{|A|^{2}-1}\cdot2^{-n\cdot\frac{\delta_{1}^{2}}{8\left(\log\left(|B|+3\right)\right)^{2}}}\ .
\end{align}
\end{lemma}

\begin{proof}
In the proof of Lemma~\ref{main_2}, we can choose the parameter $\delta_{n}$ as $\delta_{n}=\frac{1}{2}\cdot2^{-n\cdot\frac{\delta_{1}^{2}}{8\left(\log\left(|B|+3\right)\right)^{2}}}$. By~\eqref{selection} this leads to a total error rate of
\begin{align}
\alpha_{n}=(n+1)^{|A|^{2}-1}\cdot2^{-n\cdot\frac{\delta_{1}^{2}}{8\left(\log\left(|B|+3\right)\right)^{2}}}
\end{align}
for the asymptotic channel simulation, and by~\eqref{concluding} the entanglement cost for this is upper bounded by
\begin{align}
n\cdot\min_{\{M_{A\rightarrow B}^{k}\}}\max_{j}H(B|R)_{\omega^{j}}+\sqrt{n}\cdot\log\left(|B|+3\right)\cdot\sqrt{\log\left(\frac{16}{\delta_{n}^{2}}\right)}+2\cdot\log(n+1)\cdot\left(|A|^{2}-1\right)\ .
\end{align}
Since
\begin{align}
\lim_{n\rightarrow\infty}\frac{1}{n}\cdot\left(\sqrt{n}\cdot\log\left(|B|+3\right)\cdot\sqrt{\log\left(\frac{16}{\delta_{n}^{2}}\right)}\right)=\delta_{1}\ ,
\end{align}
we get an entanglement cost of $E_{C}+\delta_{1}$ (by considering the rest of the proof of the direct part of Theorem~\ref{main_1}, that is, Lemma~\ref{lem:nonblocking} and Proposition~\ref{final}).
\end{proof}

Using this lemma, we can now finally prove the following upper bound on the strong converse quantum capacity. The main idea of our proof is argue by contraction: we show that if we were able to send quantum information at a rate exceeding $E_{C}$, then we could effectively send information through a perfect channel at a higher rate than is allowed by Corollary~\ref{cor:strong}. Since our upper bound holds for any classical communication assistance, we henceforth only talk about $Q_{\leftrightarrow}$.

\begin{theorem}\label{thm:strongconverse}
Let $\cE:\cL(\cH_{A})\rightarrow\cL(\cH_{B})$ be a CPTP map and $\delta_{2}>\delta_{1}>0$. Then for every sequence of two-way classical communication assisted $\eps_{n}$-error codes for $\cE^{\otimes n}$ with asymptotic rate $R=E_{C}(\cE)+\delta_{2}$, we have
\begin{align}
\eps_{n}\geq1-(n+1)^{|A|^{2}-1}\cdot2^{-n\cdot\frac{\delta_{1}^{2}}{8\left(\log\left(|B|+3\right)\right)^{2}}}-2^{-n\cdot\frac{\delta_{2}-\delta_{1}}{E_{C}(\cE)+\delta_{1}}-1}=1-2^{-O(n)}\ .
\end{align}
\end{theorem}

\begin{proof}
We start with the perfect qubit identity channel $\cI_{2}$ and do a channel simulation for $\cE$ as defined in Definition~\ref{def:main}. As we have just seen this can be done for an entanglement cost $E_{C}(\cE)+\delta_{1}$ and an exponentially small error $\alpha_{n}=(n+1)^{|A|^{2}-1}\cdot2^{-n\cdot\frac{\delta_{1}^{2}}{8\left(\log\left(|B|+3\right)\right)^{2}}}$ (Lemma~\ref{prop:error}). Now suppose that there existed a hypothetical asymptotic code for $\cE$ allowing us to send information at a rate $R=E_{C}+\delta_{2}$ for an error rate $\eps_{n}\geq0$. Hence, in total, we would have an asymptotic code for $\cI_{2}$ at a rate $\frac{E_{C}(\cE)+\delta_{2}}{E_{C}(\cE)+\delta_{1}}>1$ for some error rate $\gamma_{n}>0$. But by the triangle inequality of the metric induced by the diamond norm and Corollary~\ref{cor:strong}, we know that
\begin{align}
(n+1)^{|A|^{2}-1}\cdot2^{-n\cdot\frac{\delta_{1}^{2}}{8\left(\log\left(|B|+3\right)\right)^{2}}}+\eps_{n}\geq\gamma_{n}\geq1-\frac{1}{2}\cdot2^{-n\cdot\left(\frac{E_{C}(\cE)+\delta_{2}}{E_{C}(\cE)+\delta_{1}}-1\right)}\ ,
\end{align}
and thus we are done.
\end{proof}

\begin{figure}[ht]
\begin{minipage}[b]{1\linewidth}
\centering
\includegraphics[width=0.6\textwidth]{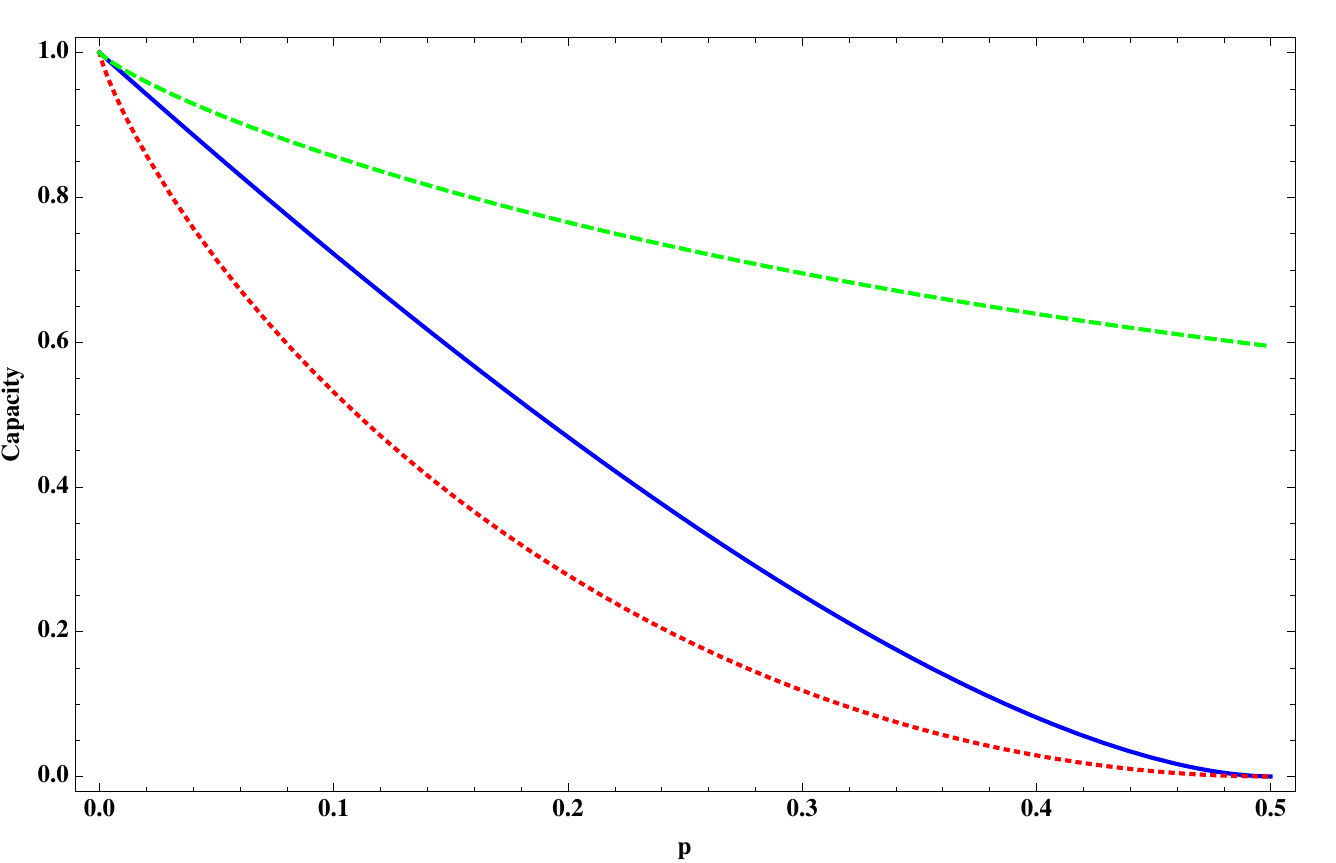}
\caption{The qubit dephasing channel with dephasing parameter $p$ - quantum capacity $Q$ (dotted line) vs. upper bound $E^{1}_{C}$ on the entanglement cost (solid line) vs. entanglement assisted quantum capacity $Q_{E}$ (dashed line).}
\label{fig:example}
\end{minipage}
\end{figure}

As an easy example, we consider the qubit erasure channel $\cE_{\mathrm{eras}}(\rho)=(1-p)\rho+p\cdot\proj{e}$ with $p\in[0,1]$. We immediately have $E_{C}(\cE_{\mathrm{eras}})\geq1-p$, and calculate~\cite{Wilde12}
\begin{align}
E_{C}(\cE_{\mathrm{eras}})\leq\max_{\psi}E_{F}((\cE_{\mathrm{eras}}\otimes\cI)(\psi))\leq E_{F}((\cE_{\mathrm{eras}}\otimes\cI)(\phi))&\leq E_{F}((1-p)\phi+p\cdot\proj{e}\otimes\frac{\1}{2})\\
&\leq (1-p)\cdot E_{F}(\phi)+p\cdot E_{F}(\proj{e}\otimes\frac{\1}{2})\\
&=1-p\ ,
\end{align}
where $\Phi$ denotes the maximally entangled state, and we used the non-regularized converse for the entanglement cost (Corollary~\ref{cor:non}), as well as the convexity of the entanglement of formation~\cite{Bennett96}. Hence $E_{C}(\cE_{\mathrm{eras}})=1-p$, and since it is also known that $Q_{\leftrightarrow}(\cE_{\mathrm{eras}})=1-p$~\cite{PhysRevLett.78.3217}, we get by Theorem~\ref{thm:strongconverse} that $Q_{\leftrightarrow}(\cE_{\mathrm{eras}})$ is a strong converse capacity. Note that, this argument for the qubit erasure channel was basically already present in~\cite{PhysRevLett.78.3217}. For generic quantum channels, we expect that the upper bound given by the entanglement cost is far from being tight. We can compare the quantum capacities of qubit channels with our upper bound from~\eqref{eq:secLimits}
\begin{align}
E_{C}(\cE_{A\rightarrow B})\leq E^{1}_{C}(\cE_{A\rightarrow B})=h\left(\frac{1}{2}+\frac{1}{2}\cdot\sqrt{1-C^{2}\left(\left(\cE_{A\rightarrow B}\otimes\cI_{A'}\right)\left(\phi_{AA'}\right)\right)}\right)\ ,
\end{align}
where $h(\cdot)$ denotes the binary Shannon entropy, and $C(\cdot)$ is defined as in~\eqref{eq:Cexpr}. For $Q_{\rightarrow}(\cE)$ this can e.g.~be evaluated for all degradable qubit channels~\cite{shor05,yard08}. As an example we mention the qubit dephasing channel $\cE_{\mathrm{deph}}(\rho)=(1-p)\rho+p\cdot\sigma_{z}\rho\sigma_{z}$ with $\sigma_{z}=\begin{pmatrix} 1&0\\ 0&-1\end{pmatrix}$, for which we get
\begin{align}
Q_{\rightarrow}(\cE_{\mathrm{deph}})=1-h(p)\leq h\left(\frac{1}{2}+\sqrt{p(1-p)}\right)=E_{C}^{1}(\cE_{\mathrm{deph}})\leq1-\frac{1}{2}\cdot h(\frac{p}{2})=Q_{E}(\cE_{\mathrm{deph}})\ .
\end{align}
where $h(\cdot)$ denotes the binary Shannon entropy~\cite{Wilde11}. As shown in Fig.~\ref{fig:example}, this is far from being tight.  However, since $Q_{\leftrightarrow}$ (and also $Q_{\leftarrow}$) can be much larger than $Q_{\rightarrow}$, and since not too much is known about these capacities, the following upper bound might be useful. We have for every qubit channel $\cE_{A\rightarrow B}$ that
\begin{align}
Q_{\leftrightarrow}(\cE_{A\rightarrow B})\leq h\left(\frac{1}{2}+\frac{1}{2}\cdot\sqrt{1-C^{2}\left(\left(\cE_{A\rightarrow B}\otimes\cI_{A'}\right)\left(\phi_{AA'}\right)\right)}\right)\ .
\end{align}


\section{Discussion and Outlook}\label{sec:dis}

We calculated the rate of entanglement needed in order to asymptotically simulate a quantum channel when classical communication is for free. Because of the free classical communication, the problem is equivalent to the question about the rate of quantum communication needed in order to simulate a quantum channel. A natural subsequent question is to ask what rate of classical communication is actually needed. However, in the spirit of general quantum channel simulations, we might even want to ask more generally about rate triples $(q,e,c)$ needed in order to achieve the channel simulation. Here $q$ denotes quantum communication, $e$ entanglement, and $c$ classical communication. The quantum reverse Shannon theorem can then be understood as e.g.~$(Q_{E},\infty,0)$ or $(0,\infty,C_{E})$, whereas our entanglement cost corresponds to e.g.~$(0,E_{C},\infty)$ or $(E_{C},0,\infty)$. Some more examples are discussed in~\cite[Figure 2]{Bennett06} and a particularly interesting case is the following. For $e=0$, $c=0$, and product state inputs, the channel simulation can be done for~\cite[Theorem 3]{Bennett06}
\begin{align}
q=\lim_{n\rightarrow\infty}\frac{1}{n}E_{P}\left(\left(\cE_{A\rightarrow B}\otimes\cI_{A'}\right)\left(\phi_{AA'}\right)^{\otimes n}\right)
\end{align}
with $\phi_{AA'}$ the maximally entangled state, and $E_{P}$ the entanglement of purification~\cite{Terhal02}
\begin{align}
E_{P}(\rho_{AB})=\min_{\rho_{AA'BB'}:\tr_{A'B'}\left[\proj{\rho}_{AA'BB'}\right]=\rho_{AB}}E_{F}(\rho_{AA'BB'})\ .
\end{align}
Now one could hope to generalize this to a channel simulation for general input states using the techniques presented above, leading to
\begin{align}
q=\lim_{n\rightarrow\infty}\frac{1}{n}\max_{\psi^{n}_{AA'}}E_{P}\left(\left(\cE^{\otimes n}_{A\rightarrow B}\otimes\cI_{A'}\right)\left(\psi^{n}_{AA'}\right)\right)\ ,
\end{align}
where $\psi^{n}_{AA'}\in\cV(\cH_{A}^{\otimes n}\otimes\cH_{A'}^{\otimes n})$, and $\cH_{A'}\cong\cH_{A}$. However, this does not work for same reason as the quantum reverse Shannon theorem can not be proven for general input states using only maximally entangled states; an issue known as entanglement spread~\cite{Bennett06,Harrow09,Hayden03,Arrow04}.\\

Another interesting question concerns the relation of $E_C({\cal E})$ and $Q_{\rightarrow}({\cal E})$. We know that $E_C({\cal E}) \geq Q_{\leftrightarrow}({\cal E})$, with the inequality typically being strict. Can we obtain a characterization of channels for which $E_C({\cal E}) = Q_{\leftrightarrow}({\cal E})$? This is an analog of the problem of characterizing bipartite states for which the distillable entanglement is equal the entanglement cost, which is still wide open.\\

\textit{Note added.} After completion of this work, security in the noisy storage model was linked to the strong converse quantum capacity of the adversary's storage device~\cite{berta11}. This means that our bound on the strong converse from Section~\ref{sec:strongconv} can also be applied directly to calculate rates for security. However, our arguments from Section~\ref{sec:noisy} apply to virtually any form of the noisy storage model, whereas the results from~\cite{berta11} are only applicable for the so-called six-state encoding. 


\section*{Acknowledgments}

We acknowledge discussions with Jonathan Oppenheim, Michael Walter, Reinhard R.~Werner, Mark M.~Wilde, and Andreas Winter. We thank Cupjin (Jiachen) Huang for pointing out an error in the proof of Lemma~\ref{main_2} in an earlier version of this manuscript and Mark M.~Wilde for pointing out an error in Figure~\ref{fig:example} in an earlier version of this manuscript. MB and MC are supported by the Swiss National Science Foundation (grant PP00P2-128455), the German Science Foundation (grants CH 843/1-1 and CH 843/2-1), and the National Centre of Competence in Research 'Quantum Science and Technology'. FB and SW are supported by the National Research Foundation and the Ministry of Education, Singapore. FB is supported by a "Conhecimento Novo" fellowship from the Brazilian agency Funda\c{c}\~ao de Amparo a Pesquisa do Estado de Minas Gerais (FAPEMIG). MB, FB and SW would like to thank the Institute Mittag-Leffler (Djursholm, Sweden), where part of this work was done.


\appendix

\section{Properties of Smooth Entropy Measures}\label{app:smooth}

\begin{lemma}\cite[Lemma 10]{datta-2008-2}
Let $\rho_{AB}\in\cP(\cH_{AB})$. Then,
\begin{align}
H_{0}(A|B)_{\rho}\geq H(A|B)_{\rho}\ .
\end{align}
\label{max}
\end{lemma}

\begin{lemma}\label{lem:logn}
Let $\rho_{A}=\sum_{j=1}^{N}p_{j}\rho_{A}^{j}\in\cP(\cH_{A})$ with $\rho_{A}^{j}\in\cP(\cH_{A})$ for $j=1,\ldots,N$. Then,
\begin{align}
H_{0}(A)_{\rho}\leq\max_{j}H_{0}(A)_{\rho^{j}}+\log N\ .
\end{align}
\end{lemma}

\begin{proof}
We have $\mathrm{rank}(M+N)\leq\mathrm{rank}(M)+\mathrm{rank}(N)$ for $M,N\in\cL(\cH_{A})$~\cite[Proposition 0.4.5]{Horn85} and hence
\begin{align}
H_{0}(A)_{\rho}&=\log\left(\mathrm{rank}\left(\sum_{j=1}^{N}p_{j}\rho_{A}^{j}\right)\right)\leq\log\left(\sum_{j=1}^{N}\mathrm{rank}\left(p_{j}\rho_{A}^{j}\right)\right)=\log\left(\sum_{j=1}^{N}\mathrm{rank}\left(\rho_{A}^{j}\right)\right)\leq\log\left(N\cdot\max_{j}\mathrm{rank}\left(\rho_{A}^{j}\right)\right)\\
&=\max_{j}H_{0}(A)_{\rho^{j}}+\log N\ .
\end{align}
\end{proof}

\begin{lemma}\label{smoothi}
Let $\eps\geq0$, $\rho_{AB}=\sum_{j=1}^{N}p_{j}\rho_{AB}^{j}\in\cS(\cH_{AB})$, $p_{j}>0$ for $j=1,\ldots,N$, and $\rho_{AB}^{j}=\sum_{k\in K}\rho_{A}^{j,k}\otimes\proj{k}_{B}$ with $\rho_{A}^{j,k}\in\cP(\cH_{AB})$ and the $\ket{k}_{B}$ mutually orthogonal. Then,
\begin{align}
H_{0}^{\eps}(A|B)_{\rho}\leq\max_{j}H_{0}^{\eps}(A|B)_{\rho^{j}}+\log N\ .
\end{align}
\end{lemma}

\begin{proof}
Let $\bar{\rho}_{AB}^{j}\in\cB^{\eps}_{qc}(\rho_{AB}^{j})$ such that $H_{0}^{\eps}(A|B)_{\rho^{j}}=H_{0}(A|B)_{\bar{\rho}^{j}}$ for each $j=1,\ldots,N$. Now define $\bar{\rho}_{AB}=\sum_{j=1}^{N}p_{j}\bar{\rho}_{AB}^{j}$ and note that $\bar{\rho}_{AB}\in\cB^{\eps}_{qc}(\rho_{AB})$. Using the definition of the alternative smooth conditional max-entropy and its form on quantum-classical states (Lemma~\ref{lem:h0class}), it follows that
\begin{align}
H_{0}^{\eps}(A|B)_{\rho}\leq H_{0}(A|B)_{\bar{\rho}}=\max_{k\in K}H_{0}(A)_{\sum_{j}p_{j}\bar{\rho}^{j,k}}\ .
\end{align}
Using the preceding lemma (Lemma~\ref{lem:logn}) and again the structure of quantum-classical states (Lemma~\ref{lem:h0class}), we conclude
\begin{align}
\max_{k\in K}H_{0}(A)_{\sum_{j}p_{j}\bar{\rho}_{A}^{j,k}}\leq\max_{k\in K}\max_{j}H_{0}(A)_{\bar{\rho}^{j,k}}+\log N=\max_{j}H_{0}(A|B)_{\bar{\rho}^{j}}+\log N=\max_{j}H_{0}^{\eps}(A|B)_{\rho^{j}}+\log N\ .
\end{align}
\end{proof}

\begin{lemma}\label{lem:smoothing}
Let $\eps>0$, $\rho_{AB}=\sum_{k}\rho_{A}^{k}\otimes\proj{k}_{B}\in\cS(\cH_{AB})$, and the $\ket{k}_{B}$ mutually orthogonal. Then, the smoothing in $H_{0}^{\eps}(A|B)_{\rho}$ can without lost of generality be restricted to states that commute with $\rho_{AB}$.
\end{lemma}

\begin{proof}
The crucial step is to see that for every $\sigma_{AB}=\sum_{k}\sigma_{A}^{k}\otimes\proj{k}_{B}\in\cB^{\eps}_{qc}(\rho_{AB})$, there exists a unitary $U_{AB}=\sum_{k}U_{A}^{k}\otimes\proj{k}_{B}$ such that $U_{AB}\sigma_{AB}U_{AB}^{\dagger}\in\cB^{\eps}_{qc}(\rho_{AB})$ and $[U_{AB}\sigma_{AB}U_{AB}^{\dagger},\rho_{AB}]=0$. For this, just choose $U_{A}^{k}$ to be the unitary that maps the eigenbasis of $\sigma_{A}^{k}$ to the eigenbasis of $\rho_{A}^{k}$. Therefore $[U_{AB}\sigma_{AB}U_{AB}^{\dagger},\rho_{AB}]=0$, and furthermore by Lemma~\ref{lem:eigenvalue}
\begin{align}
\eps\geq\|\rho_{AB}-\sigma_{AB}\|_{1}=\sum_{k}\|\rho_{A}^{k}-\sigma_{A}^{k}\|_{1}\geq\sum_{k}\|P_{A}^{k}-Q_{A}^{k}\|_{1}=\sum_{k}\|\rho_{A}^{k}-U_{A}^{k}\sigma_{A}^{k}(U_{A}^{k})^{\dagger}\|_{1}=\|\rho_{AB}-U_{AB}\sigma_{AB}U_{AB}^{\dagger}\|_{1}\ ,
\end{align}
where $P_{A}^{k},Q_{A}^{k}$ denote the eigenvalue distributions of $\rho_{A}^{k},\sigma_{A}^{k}$ respectively.
\end{proof}

The definition of the smooth alternative conditional max-entropy can be specialized canonically to classical probability distributions.

\begin{definition}\label{def:classical}
Let $\eps\geq0$, $X$ and $Y$ be random variables with range $\cX$ and $\cY$ respectively, and joint probability distribution $P_{XY}$. The \textit{max-entropy} of $X$ conditioned on $Y$ is defined as
\begin{align}
H_{0}(X|Y)_{P}=\max_{y\in\cY}\log|\mathrm{supp}\left(P_{X}^{y}\right)|\ ,
\end{align}
where $P_{X}^{y}$ denotes the function $P_{X}^{y}:x\mapsto P_{XY}(x,y)$. The \textit{smooth max-entropy} of $X$ conditioned on $Y$ is defined as
\begin{align}
H_{0}^{\eps}(X|Y)_{P}=\inf_{\bar{P}_{XY}\in\cB^{\eps}_{c}(P_{XY})}H_{0}(X|Y)_{\bar{P}}\ ,
\end{align}
where $\cB^{\eps}_{c}(P_{XY})$ denotes the set of non-negative linear functions $\bar{P}_{XY}:\cX\times\cY\rightarrow\mathbb{R}^{+}$ such that $\|P_{XY}-\bar{P}_{XY}\|_{1}\leq\eps$.
\end{definition}

The following is an entropic formulation of the classical asymptotic equipartition property.

\begin{lemma}\cite[Theorem 1]{Holenstein11}\label{lem:classicalaep}
Let $X$ and $Y$ be random variables with range $\cX$ and $\cY$ respectively, and joint probability distribution $P_{XY}$. Furthermore let $\eps>0$, $n\geq1$, and let $P_{X^{n}Y^{n}}^{n}= P_{X_{1}Y_{1}}\times\ldots\times P_{X_{n}Y_{n}}$ be the $n$-fold product probability distribution over $\cX^{n}\times\cY^{n}$. Then
\begin{align}
\frac{1}{n}H_{0}^{\eps}(X^{n}|Y^{n})_{P^{n}}\leq H(X|Y)_{P}+\frac{\log\left(|X|+3\right)\cdot\sqrt{\log\left(\frac{1}{\eps^{2}}\right)}}{\sqrt{n}}\ .
\end{align}
\end{lemma}

This can be generalized to the following quantum-classical asymptotic equipartition property.

\begin{lemma}\label{lem:aep}
Let $\eps>0$, $n\geq1$, $\rho_{AB}=\sum_{k}\rho_{A}^{k}\otimes\proj{k}_{B}\in\cS(\cH_{AB})$ and the $\ket{k}_{B}$ mutually orthogonal. Then,
\begin{align}
\frac{1}{n}H_{0}^{\eps}(A|B)_{\rho^{\otimes n}}\leq H(A|B)_{\rho}+\frac{\log\left(|A|+3\right)\cdot\sqrt{\log\left(\frac{1}{\eps^{2}}\right)}}{\sqrt{n}}\ .
\end{align}
\end{lemma}

\begin{proof}
The basic idea is that by Lemma~\ref{lem:smoothing}, the smoothing of the alternative conditional max-entropy can be restricted to states that commute with the initial state, and hence all states that appear are diagonal in the same basis. Working in this basis, this then allows us to use the classical asymptotic equipartition property (Lemma~\ref{lem:classicalaep}). In more detail, we calculate
\begin{align}
\frac{1}{n}H_{0}^{\eps}(A|B)_{\rho^{\otimes n}}=\frac{1}{n}\min_{\bar{\rho}^{n}_{AB}\in\cB^{\eps}_{qc}(\rho_{AB}^{\otimes n})}H_{0}(A|B)_{\bar{\rho}^{n}}=\frac{1}{n}\min_{\bar{P}_{AB}^{n}\in\cB^{\eps}_{c}(P_{AB}^{n})}H_{0}(A|B)_{\bar{P}^{n}}\ ,
\end{align}
where the second equality is due to Lemma~\ref{lem:smoothing}, $P_{AB}^{n}$ is the eigenvalue distribution of $\rho_{AB}^{\otimes n}$, and $\cB^{\eps}_{c}(\cdot)$ is defined as in Definition~\ref{def:classical}. Moreover,  we conclude by the definition of the classical smooth conditional max-entropy (Definition~\ref{def:classical}), and the classical asymptotic equipartition property (Lemma~\ref{lem:classicalaep})
\begin{align}
\frac{1}{n}\min_{\bar{P}_{AB}^{n}\in\cB^{\eps}_{c}(P_{AB}^{n})}H_{0}(A|B)_{\bar{P}^{n}}=\frac{1}{n}H_{0}^{\eps}(A|B)_{P^{n}}&\leq H(A|B)_{P}+\frac{\log\left(|A|+3\right)\cdot\sqrt{\log\left(\frac{1}{\eps^{2}}\right)}}{\sqrt{n}}\\
&=H(A|B)_{\rho}+\frac{\log\left(|A|+3\right)\cdot\sqrt{\log\left(\frac{1}{\eps^{2}}\right)}}{\sqrt{n}}\ ,
\end{align}
where $P_{AB}$ denotes the eigenvalue distribution of $\rho_{AB}$.
\end{proof}


\section{The Post-Selection Technique}

The following proposition lies at the heart of the \textit{post-selection technique}.

\begin{proposition}\cite{ChristKoenRennerPostSelect}
Let $\eps>0$ and $\cE^{n}_{A}$ and $\cF^{n}_{A}$ be CPTP maps from $\cL(\cH_{A}^{\otimes n})$ to $\cL(\cH_{B})$. If there exists a CPTP map $K_{\pi}$ for any permutation $\pi$ such that $(\cE^{n}_{A}-\cF^{n}_{A})\circ\pi=K_{\pi}\circ(\cE^{n}_{A}-\cF^{n}_{A})$, then $\cE^{n}_{A}$ and $\cF^{n}_{A}$ are $\eps$-close whenever
\begin{align}
\left\|((\cE^{n}_{A}-\cF^{n}_{A})\otimes\cI_{RR'})(\zeta^{n}_{ARR'})\right\|_{1}\leq\eps(n+1)^{-(|A|^{2}-1)}\ ,
\end{align}
where $\zeta^{n}_{ARR'}$ is a purification of the de Finetti state $\zeta_{AR}^{n}=\int\sigma_{AR}^{\otimes n}d(\sigma_{AR})$ with $\sigma_{AR}\in\cV(\cH_{A}\otimes\cH_{R})$, $\cH_{A}\cong\cH_{R}$ and $d(\cdot)$ the measure on the normalized pure states on $\cH_{A}\otimes\cH_{R}$ induced by the Haar measure on the unitary group acting on $\cH_{A}\otimes\cH_{R}$, normalized to $\int d(\cdot)=1$. Furthermore we can assume without loss of generality that $|R'|\leq(n+1)^{|A|^{2}-1}$.
\label{posti}
\end{proposition}

\begin{lemma}\cite[Corollary D.6]{Berta09_2}
Let $\zeta_{AR}^{n}=\int\sigma_{AR}^{\otimes n}d(\sigma_{AR})$ as in Proposition~\ref{posti}. Then $\zeta_{AR}^{n}=\sum_{i}p_{i}\left(\omega^{i}_{AR}\right)^{\otimes n}$ with $\omega^{i}_{AR}\in\cV(\cH_{A}\otimes\cH_{R})$, $i\in\{1,2,\ldots,(n+1)^{2|A||R|-2}\}$, and $\{p_{i}\}$ a probability distribution.
\label{mario}
\end{lemma}


\section{Technical Lemmas}\label{app:tech}

\begin{lemma}\cite[Lemma 6]{Tomamichel09}\label{a:1}
Let $\rho$, $\sigma\in\cS_{\leq}(\cH)$. Then
\begin{align}
\frac{1}{2}\cdot\|\rho-\sigma\|_{1}\leq P(\rho,\sigma)\leq\sqrt{\|\rho-\sigma\|_{1}+|\tr[\rho]-\tr[\sigma]|}\ .
\end{align}
\end{lemma}

\begin{lemma}\label{lem:12norm}\cite{Horn85}
Let $M\in\mathbb{C}^{a\times b}$ for $a,b\in\mathbb{N}$. Then $\|M\|_{2}\leq\|M\|_{1}\leq\sqrt{\mathrm{rank}(M)}\cdot\|M\|_{2}$.
\end{lemma}

\begin{lemma}\cite[Section 5.2]{MeyerBook}\label{lem:sub}
Let $M\in\mathbb{C}^{a\times b}$ and $N\in\mathbb{C}^{b\times c}$ for $a,b,c\in\mathbb{N}$. Then $\|M\cdot N\|_{2}\leq\|M\|_{2}\|N\|_{2}$.
\end{lemma}

\begin{lemma}\label{lem:net}
Let $0<\eps<1$ and $D,d>0$. Furthermore let $\cN_{D}^{d}=\left\{w\in\mathbb{C}^{d}\mid\|w\|_{2}\leq D\right\}$ and let $\cT$ be some subset of $\cN_{D}^{d}$. Then, there exists a subset $\cT_{\eps}\subseteq\cT$ with $|\cT_{\eps}|\leq\left(\frac{2D}{\eps}+1\right)^{2d}$, such that for every vector $v\in\cT$, there exists a vector $v_{\eps}\in\cT_{\eps}$ with $\|v-v_{\eps}\|_{2}\leq\eps$.
\end{lemma}

\begin{proof}
The proof is inspired by~\cite[Lemma II.4]{Leung04}. Let $\cT_{\eps}=\{v_{i}\}_{i=1,\ldots,m}$ be a maximal subset of $v\in\cT$ satisfying $\|v_{i}-v_{j}\|_{2}\geq\eps$ for all $i,j$.\footnote{Such a subset can be constructed by starting with an arbitrary vector $v_{1}\in\cT$, as a next step taking another vector $v_{2}\in\cT$ with $\|v_{1}-v_{2}\|_{2}\geq\eps$, and then $v_{3}\in\cT$ with $\|v_{1}-v_{3}\|_{2}\geq\eps$, $\|v_{2}-v_{3}\|_{2}\geq\eps$ etc. A subset constructed like this becomes maximal as soon as it is not possible to add another vector $v_{k}\in\cT$, such that $\|v_{k}-v_{i}\|_{2}\geq\eps$ for all vectors $v_{i}$ that are already in the subset.} It remains to estimate $m$. As subsets of $\mathbb{R}^{2d}$, the open balls of radius $\eps/2$ about each $v_{i}\in\cT_{\eps}$ are pairwise disjoint, and all contained in the ball of radius $D+\eps/2$ centered at the origin. Hence 
\begin{align}
m\cdot\left(\eps/2\right)^{2d}\leq\left(D+\eps/2\right)^{2d}\ .
\end{align}
\end{proof}

\begin{lemma}\cite[Corollary 3.3]{Sion58}\label{lem:minimax}
Let $X$ and $Y$ be convex, compact sets and $f$ a real valued function on $X\times Y$, that is convex in the first argument, concave in the second argument and continuous in both. Then,
\begin{align}
\inf_{x\in X}\sup_{y\in Y}f(x,y)=\sup_{y\in Y}\inf_{x\in X}f(x,y)\ .
\end{align}
\end{lemma}

\begin{lemma}\cite{NieChu00Book}\label{lem:eigenvalue}
Let $\rho,\sigma\in\cP(\cH)$, and denote the corresponding eigenvalue distribution by $P_{X},Q_{X}$ respectively. Then,
\begin{align}
\|\rho-\sigma\|_{1}\geq\|P_{X}-Q_{X}\|_{1}\ .
\end{align}
\end{lemma}

\begin{lemma}\cite[Theorem 1]{Audenaert07}\label{lem:neumann}
Let $\rho_{A},\sigma_{A}\in\cS(\cH_{A})$ with $\rho_{A}\approx_{\eps}\sigma_{A}$ for some $\eps\geq0$. Then,
\begin{align}
|H(A)_{\rho}-H(A)_{\sigma})|\leq\eps\cdot\log(|A|-1)+h(\eps)\ ,
\end{align}
where $h(\cdot)$ denotes the binary Shannon entropy.
\end{lemma}

\begin{lemma}\cite{Alick04}\label{lem:fannes}
Let $\rho_{AB},\sigma_{AB}\in\cS(\cH_{AB})$ with $\|\rho_{AB}-\sigma_{AB}\|_{1}\leq\eps$ for some $\eps\geq0$. Then,
\begin{align}
|H(A|B)_{\rho}-H(A|B)_{\sigma}|\leq4\eps\cdot\log|A|+2h(\eps)\ ,
\end{align}
where $h(\cdot)$ denotes the binary Shannon entropy.
\end{lemma}

\begin{lemma}\label{lem:nielsen}
Let $\rho_{AB},\sigma_{AB}\in\cS(\cH_{AB})$ with $\rho_{AB}\approx_{\eps}\sigma_{AB}$ for some $\eps\geq0$. Then,
\begin{align}
|E_{F}(\rho_{AB})-E_{F}(\sigma_{AB})|\leq8\eps\cdot\log|A|+2h(2\eps)\ ,
\end{align}
where $h(\cdot)$ denotes the binary Shannon entropy.
\end{lemma}

\begin{proof}
The proof is the same as the original one~\cite{Nielsen00}, but uses the (improved) continuity of the conditional von Neumann entropy (Lemma~\ref{lem:fannes}) instead of the continuity of the unconditional von Neumann entropy (Lemma~\ref{lem:neumann}).
\end{proof}

\begin{lemma}\cite[Lemma 1]{Uhlmann98}\label{lem:uhlman}
Let $\rho_{AB}\in\cS(\cH_{AB})$. Then the minimization over all pure states decompositions $\rho_{AB}=\sum_{i}p_{i}\rho_{AB}^{i}$ in the entanglement of formation $E_{F}(\rho_{AB})=\min_{\{p_{i},\rho^{i}\}}\sum_{i}p_{i}H(A)_{\rho^{i}}$ (Definition~\ref{def:formation}), is taken for a decomposition with at least $\mathrm{rank}(\rho_{AB})$ and at most $\mathrm{rank}(\rho_{AB})^{2}$ elements.
\end{lemma}

\begin{lemma}\cite[Proposition 4.3]{werner04}\label{lem:werner}
Let $\cE:\cL(\cH_{A})\mapsto\cL(\cH_{B})$ be a quantum channel. Then,
\begin{align}
1-\min_{\rho\in\cS(\cH_{A})}F_{e}(\rho,\cE)\leq4\sqrt{1-F_{c}(\cE)}\leq4\sqrt{\|\cE-\cI\|_{\diamond}}\leq8\left(1-\min_{\rho\in\cS(\cH_{A})}F_{e}(\rho,\cE)\right)^{1/4}\ ,
\end{align}
where $F_{c}(\cE)=\bra{\phi}(\cE\otimes\cI)(\phi)\ket{\phi}$ with $\phi_{AA'}$ the maximally entangled state on $\cH_{A}\otimes\cH_{A'}$, and $F_{e}(\rho,\cE)=\bra{\rho}(\cE\otimes\cI)(\rho)\ket{\rho}$ with $\rho_{AA'}\in\cV(\cH_{A}\otimes\cH_{A'})$ a purification of $\rho_{A}$.
\end{lemma}


\bibliography{ECQC_2}

\end{document}